\documentclass[1p,times,preprint]{elsarticle}
\usepackage[english]{babel}
\usepackage{tensor}
\usepackage{graphicx}
\usepackage{amsmath}
\usepackage{amssymb}
\usepackage{amsfonts}
\usepackage{dcolumn}
\usepackage{bm}
\usepackage{xcolor}
\usepackage{ulem}
\usepackage{tikz}
\usepackage{subcaption}
\usepackage{comment}
\usepackage{verbatim}
\usepackage{fancyvrb}
\usepackage{cancel}
\usepackage{multirow}
\usepackage{lscape}
\usepackage{txfonts}
\usepackage{mathtools}
\usepackage{soul}
\usepackage{url}
\usepackage{makecell}
\usepackage[pdftex]{pict2e}
\usepackage{microtype}

\def\bra<#1|{\mathinner{\langle\,{#1}\,\vert}} 
\def\ket|#1>{\mathinner{\vert\,{#1}\,\rangle}} 
\def\red|#1|{\mathinner{\!\vert\,{#1}\,\vert\!}}
\def\braket<#1>{\mathinner{\langle\,{#1}\,\rangle}} 

\def\redmem#1#2#3{  \left\langle #1 \left\Vert  
                  #2 \right\Vert #3 \right\rangle   }

\begin{document}

\title{Second-order Rayleigh-Schr\"odinger Perturbation Theory for the {\sc Grasp}2018 Package: Core-valence Correlations}

\author[TFAI]{G. Gaigalas}
\address[TFAI]{Institute of Theoretical Physics and Astronomy, 
               Vilnius University, Saul\.{e}tekio Ave. 3, LT-10257 Vilnius, Lithuania}
\ead{gediminas.gaigalas@tfai.vu.lt}

\author[TFAI]{P. Rynkun}
\ead{pavel.rynkun@tfai.vu.lt}

\author[TFAI]{L. Kitovien\.{e}}
\ead{laima.radziute@tfai.vu.lt}


%
%
\begin{abstract}
The General
Relativistic Atomic Structure package [{\sc Grasp}2018, C. Froese Fischer, G. Gaigalas, P. J\"onsson,
J. Biero\'n, Comput. Phys. Commun. (2019), DOI: 10.1016/j.cpc.2018.10.032],
is based on multiconfiguration Dirac-Hartree-Fock and relativistic configuration interaction (RCI) methods for energy structure calculations. Atomic state function used in the program is built from the set of configuration state functions (CSFs). The valence-valence,
core-valence and core-core correlations are explicitly included through expansions over CSFs in RCI.
We present a combination of RCI and the stationary second-order Rayleigh-Schr\"odinger many-body perturbation theory in irreducible tensorial form to account for electron core-valence correlations when an atom or ion has any number of valence electrons. This newly developed method, which offers two ways of use, allows a significant reduction of the CSF space for complex atoms and ions.
We also demonstrate how the method and program works for energy structure calculation of Cl III element.

\end{abstract}

\maketitle


\newpage

\section{Introduction}

There are many powerful theoretical methods for taking into account both relativistic and correlation effects in calculation atomic properties of
any-electron atoms and ions in our days.
This can
be done, for example, by  using various versions of many-body perturbation theory (MBPT),
the configuration interaction method (CI), the relativistic configuration interaction (RCI), the random phase approximation
with exchange (RPAE), the  multiconfiguration Hartree-Fock method (MCHF)
\cite{fischerBook:77} or the multiconfiguration Dirac-Hartree-Fock (MCDHF) method \cite{grantBook:07}.
But each of them has its own disadvantages for getting atomic data in extremely good accuracy. For example, the converging to accurate results are very slowly for MCHF, MCDHF, and CI. This leads to very large expansion of atomic state function (ASF) for complex atoms.
There are practical and theoretical difficulties with perturbation theory for
degenerate states, especially in the choice of model space~\cite{LindgrenBook:82} also. The structure of terms of the PT series often leads to one-
and two-particle operators which almost in all versions of many-body perturbation theory are not in irreducible tensorial form and which can not use advantage of Racah algebra~\cite{Gaigalas_1996,Gaigalas_1997}.

Combination of CI and MBPT methods~\cite{Bogetal:1997,DzuFla:07} is probably the most efficient and consistent way to account simultaneously
correlation and relativistic effects in such complex many-electron atoms with open $f$-shells as lanthanides and actinides for energy spectra
and other properties calculations.

Here we present such implementation of MBPT to the {\sc Grasp} code~\cite{grasp2018} in which core-valence correlations can be taken into account with the help of MBPT and rest correlations are included in ordinary way (RCI) by {\sc Grasp} package. We are using the most suitable for these systems the irreducible tensorial form of Rayleigh-Schr\"odinger stationary many-body perturbation theory~\cite{Meratal:86,Gaigalas:89}. This allow us the calculation of terms of
the perturbation series to divide into the calculation of spin-angular
terms by using Racah algebra~\cite{Gaigalas_1996,Gaigalas_1997,Gaigalas:2022} and into the accompanying radial integrals. The latter are the more
straightforward and can be handled by methods such as those of the {\sc Grasp} code
\cite{grasp2013,grasp2018}. Such packages have a modular structure, and modules for
generating the contribution of diagrams representing terms of the MBPT series in irreducible tensorial form
can easily be added on to the system.
The newly developed method, which combines the RCI and the stationary second-order Rayleigh-Schr\"odinger many-body perturbation theory in irreducible tensorial form (RCI+RSMBPT), 
and its implementation in the {\sc Grasp} package is described in detail in the sections below.

\section{General theory}

\subsection{Main theories in the {\sc Grasp} package}
\label{sec:GRASP_proc}
The MCDHF method is
based on the Dirac-Coulomb (DC) Hamiltonian~\cite{grantBook:07,Fisetal:16a}
\begin{equation}
H_{DC} = \sum_{i=1}^N \left( c \; { \bm{\alpha} }_i \cdot
                    {\bf{ p }}_i
         + (\beta_i -1)c^2 + V^N_i \right)
         + \sum_{i>j}^N \frac{1}{r_{ij}},
\end{equation}
where $V^N$ is the monopole part of the electron-nucleus Coulomb interaction, $\bm{ \alpha }$ and $\beta$ are the $4 \times 4$ Dirac matrices, and $c$ is the speed of light in atomic units.
The atomic state functions (ASFs) were obtained as linear
combinations of symmetry adapted configuration state functions (CSFs)~\cite{Jonatal:2023}
\begin{equation}
\label{ASF}
\Psi({\mathit \gamma} PJM)  = \sum_{j=1}^{N_{CSFs}} c_{j} \Phi(\gamma_{j}PJM).
\end{equation}
Here $J$ and $M$ are the angular quantum numbers and $P$ is parity.
$\gamma_j$ denotes other appropriate labeling of the configuration state function $j$,
for example orbital occupancy and coupling scheme. Normally the label ${\mathit \gamma}$ of the atomic state
function is the same as the label of the dominating CSF.
For these calculations the spin-angular approach \cite{Gaigalas_1996,Gaigalas_1997},
which is based on the second quantization in coupled
tensorial form, on the angular momentum theory in
three spaces (orbital, spin, and quasispin) and on the reduced coefficients of fractional parentage,
was used. It allows us to study configurations with open $f$-shells without any restrictions.
The CSFs are built from products of one-electron Dirac orbitals. 
Based on a weighted energy average of several states, the so-called extended optimal level (EOL) scheme \cite{grasp1989},
both the radial parts of the Dirac orbitals and the 
expansion coefficients were optimized to self-consistency in the relativistic self-consistent field procedure \cite{Fisetal:16a}.

In RCI computations, the atomic state function is expanded in CSFs, and only 
the expansion coefficients are determined by diagonalizing the Hamiltonian matrix~\cite{Jonatal:2023}.
The RCI method is also used to include the transverse-photon (Breit) interaction 
and QED corrections: vacuum polarization
and self-energy~\cite{Jonatal:2023}.
More details about MCHDF and RCI methods can be found in~\cite{Fisetal:16a,Jonatal:2023}.

\subsection{Zero-first-order method}
\label{sec:ZFOm}

One of Brillouin-Wigner perturbation theory versions is widely used by the {\sc Grasp} community. 
In this version, according to Brillouin-Wigner perturbation theory~\cite{LindgrenBook:82,Kato_2001},
the CSF space can be divided into two parts:

i) a principal part ($P$), which contains CSFs that account for the major parts of the wave functions and is referred to as a zero-order partitioning; 

ii) an orthogonal complementary part ($Q$), which contains CSFs that represent minor corrections and 
is referred to as a first-order partitioning.

Interaction between $P$ and $Q$ is assumed to be the lowest-order
perturbation. The total energy functional is partitioned into the
zero-order part ($H^{(0)}$) and the residual part ($V$). The Dirac-Fock 
energy functional is chosen as the zero-order part; the
residual part then represents a correlation energy functional.
The second-order Brillouin-Wigner perturbation theory then
leads to,
\begin{eqnarray}
(E-H^{(0)}_{QQ})^{-1}V_{QP}\Psi_P = \Psi_Q,
\nonumber \\
\left[H^{(0)}_{PP} + V_{PP} + V_{PQ}(E-H^{(0)}_{QQ})^{-1}V_{QP}\right]\Psi_P = E \Psi_P .
\end{eqnarray}
The above equations define the first-order correlation operator 
and the second-order effective Hamiltonian operator for
the {$P$-space}, respectively. In the brackets of the second equation, 
the first and second terms compose the total energy
functional in the {$P$-space}, and the third term represents the
second-order correction to the correlation energy functional
in the $P$-space. The non-linear effective Hamiltonian equation 
is written in a linearized form,
\begin{equation}
\label{Hamiltonian}
\begin{pmatrix} H^{(0)}_{PP} +V_{PP} ~~~~V_{PQ} \\ V_{QP}~~~~~~~~~~~~~~~H^{(0)}_{QQ}\end{pmatrix} 
\begin{pmatrix} \Psi_P \\ \Psi_Q \end{pmatrix}
=
E\begin{pmatrix} \Psi_P \\ \Psi_Q \end{pmatrix} .
\end{equation}
The requirement that the total energy functional ($E$) is
stationary with respect to variations in spin-orbitals ($\{\phi\}$) under 
the normalization and the orthogonality conditions leads
to a set of the Euler-Lagrange equations,
\begin{equation}
\frac{\delta E [\{\phi\}]}{\delta \phi_a} = \mu_a \phi_a + \sum_{b \neq a} \mu_{ab}\phi_b \;,
\end{equation}
where $\{\mu\}$ are the Lagrange multipliers. The above equations
are nothing but reduced MCDHF equations. That is to say, an
apparent connection between the second-order Brillouin-Wigner
perturbation energy functional and a set of reduced 
MCDHF equations is provided.

Block $H_{QQ}^{(0)}$ is diagonal in the Hamiltonian matrix (see Eq. (\ref{Hamiltonian})).
As a result, computation time and size required 
for the construction of the Hamiltonian matrix are reduced. 
This method, named as zero-first-order method (ZF), has the potential for
taking a very large configuration space into account, 
which is almost unachievable by full MCDHF and RCI methods~\cite{Jonatal:2023}, 
and for allowing accurate calculation to be performed with relatively small computational resources 
provided the $Q$-space contributes perturbatively to 
the $P$-space. 
However, all above mentioned methods (see subsections~\ref{sec:GRASP_proc} and \ref{sec:ZFOm})
still lead to a very large space of ASF and need a big amount of computational resources in case of very accurate calculations of complex atoms. 
So, the further improvements are welcome.
One of the solutions to this problem is implementation of the combination of RCI method with 
the Rayleigh-Schr\"odinger perturbation theory in irreducible tensorial form in the {\sc Grasp} using usual PT formalism.
This is what is going to be discussed in the sections below.

\subsection{Rayleigh-Schr\"odinger perturbation theory in irreducible tensorial form}
\label{sec:PT_Mano}

The Brillouin-Wigner form of perturbation theory is formally very simple. However, the effective operator of second order depends on the exact energy of considered state. This requires self-consistency procedure, and limits the application to one energy level at a time. 
There are other issues with implementation of this theory~\cite{LindgrenBook:82} in the {\sc Grasp}, too.
The Rayleigh-Schr\"odinger perturbation theory, another version of PT, does not have these shortcomings, and therefore is more suitable for many-electron calculations than the Brillouin-Wigner PT. 

There are several variations of the second-order Rayleigh-Schr\"odinger perturbation theory. 
Most of them deal with determinants instead of ASF in the form of Eq. (\ref{ASF}).
Only one of them~\cite{Meratal:86,Gaigalas:89} is formulated in irreducible tensorial form
which gives the opportunity to include core-valence correlations with any number of valence electrons.
This allows us for applying it to various applications to use
the combination~\cite{Gaigalas_1996} of the angular momentum theory~\cite{Yutetal:62a}, as
described in~\cite{JucBan:77a}, the concept of irreducible tensorial sets~\cite{FanRac:59a,Jud:67a,RudKan:84a}, a generalized
graphical approach~\cite{Gaigatal:85}, the second quantization in coupled tensorial form~\cite{RudKan:84a}, the
quasi-spin approach~\cite{Rud:97a}, and the use of reduced coefficients of fractional parentage~\cite{Gaietal:2000a}, 
the same as used by the {\sc Grasp} for 
calculation one- and two- particle operators~\cite{Gaigalas:2022}. 
Given that the spin-angular structure of terms of the PT series 
of this version leads to one-, two-, and three- particle operators in tensorial form, this version~\cite{Meratal:86,Gaigalas:89} of PT theory is ideal for the {\sc Grasp} package.

\begin{figure*}
\begin{center}
\setlength{\unitlength}{1mm}
\begin{picture}(148,32)
\thicklines
\put(10,19){\vector(0,1){10}}
\put(10,29){\vector(0,1){2}}
\put(7.5,27){\makebox(0,0)[t]{\small{$m$}}}
\put(20,19){\vector(0,1){10}}
\put(20,29){\vector(0,1){2}}
\put(22,27){\makebox(0,0)[t]{\small{$n$}}}
\multiput(10,19)(1,0){10}{\circle*{0.35}}
\put(10,09){\line(0,1){10}}
\put(10,09){\vector(0,1){3}}
\put(7.5,12){\makebox(0,0){\small{$m^{\prime}$}}}
\put(10,07){\vector(0,1){3}}
\put(20,09){\line(0,1){10}}
\put(20,09){\vector(0,1){3}}
\put(22.5,12){\makebox(0,0){\small{$n^{\prime}$}}}
\put(20,07){\vector(0,1){3}}
\put(15,04){\makebox(0,0){$ A_{1}$}}
\put(26,19){\makebox(0,0) [l] {$\displaystyle{\sim \sum_{k}\frac{1}{\sqrt{[k]}}\, \Theta^{(1)}(n_m\ell_m j_m, n_n\ell_n j_n, n_{m'}\ell_{m'} j_{m'}, n_{n'}\ell_{n'} j_{n'}, \; \Xi)}$}}
\put(120,19){\vector(0,1){12}}
\put(120,30){\oval(5,5)[b]}
\put(120,19){\circle*{1,7}}
\put(116,27){\makebox(0,0)[t]{$n_{m}j_{m}$}}
\put(130,19){\vector(0,1){12}}
\put(130,30){\oval(5,5)[b]}
\put(130,19){\circle*{1,7}}
\put(134,27){\makebox(0,0)[t]{$n_{n}j_{n}$}}
\put(120,19.6){\line(10,0){10}}
\put(120,19.3){\line(10,0){10}}
\put(120,19){\line(10,0){10}}
\put(120,18.7){\line(10,0){10}}
\put(120,18.4){\line(10,0){10}}
\put(123,17){\line(0,4){4}}
\put(128,19.6){\vector(-1,0){3}}
\put(128,18.4){\vector(-1,0){3}}
\put(125,22){\makebox(0,0){$k$}}
\put(120,09){\line(0,1){10}}
\put(115.5,14){\makebox(0,0)[t]{$n_{m'}j_{m'}$}}
\put(117.5,19){\makebox(0,0){${+}$}}
\put(120,06){\vector(0,1){4}}
\put(120,08){\oval(5,5)[t]}
\put(130,09){\line(0,1){10}}
\put(135,14){\makebox(0,0)[t]{$n_{n'}j_{n'}$}}
\put(132.7,19){\makebox(0,0){${-}$}}
\put(130,06){\vector(0,1){4}}
\put(130,08){\oval(5,5)[t]}
\put(125,04){\makebox(0,0){$ A_{2}$}}
\end{picture}
\caption{Two-electron scalar operator, when the second quantization operators
are coupled by pairs $[a^{(j)} \times \tilde{a}^{(j')}]^{(k)}$.}
\label{d-a}
\end{center}
\end{figure*}
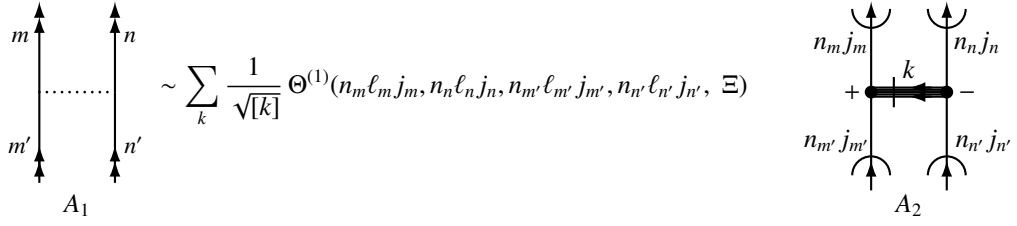

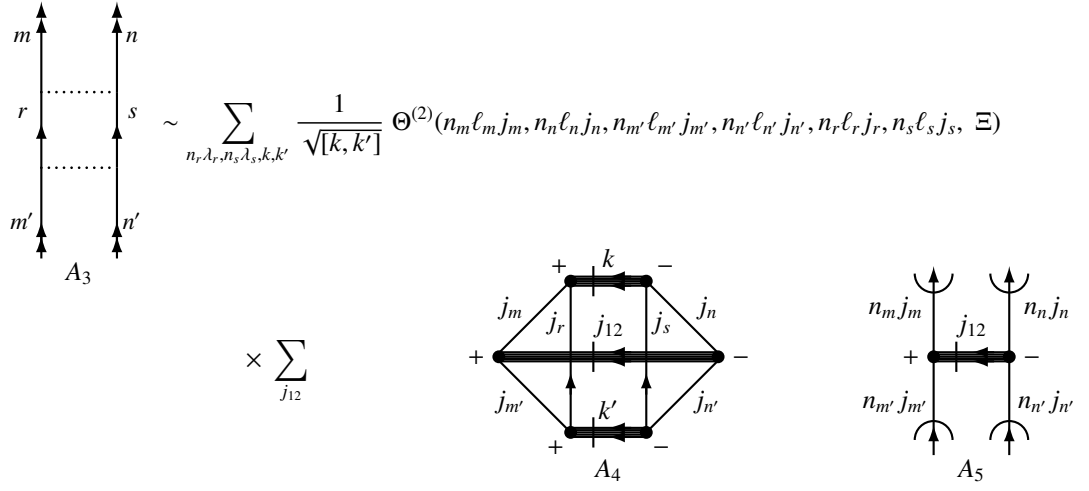
\begin{figure}
\begin{center}
\setlength{\unitlength}{1mm}
\begin{picture}(148,67)
\thicklines
\put(10,54){\vector(0,1){10}}
\put(10,64){\vector(0,1){2}}
\put(7.5,62){\makebox(0,0)[t]{\small{$m$}}}
\put(20,54){\vector(0,1){10}}
\put(20,64){\vector(0,1){2}}
\put(22,62){\makebox(0,0)[t]{\small{$n$}}}
\multiput(10,54)(1,0){10}{\circle*{0.35}}
\put(10,44){\line(0,10){10}}
\put(10,48){\vector(0,1){2}}
\put(7.5,52){\makebox(0,0)[t]{\small{$r$}}}
\put(20,44){\line(0,10){10}}
\put(20,48){\vector(0,1){2}}
\put(22,52){\makebox(0,0)[t]{\small{$s$}}}
\multiput(10,44)(1,0){10}{\circle*{0.35}}
\put(10,34){\line(0,1){10}}
\put(10,34){\vector(0,1){3}}
\put(7.5,37){\makebox(0,0){\small{$m'$}}}
\put(10,32){\vector(0,1){3}}
\put(20,34){\line(0,1){10}}
\put(20,34){\vector(0,1){3}}
\put(15,30){\makebox(0,0){$ A_{3}$}}
\put(22,37){\makebox(0,0){\small{$n'$}}}
\put(20,32){\vector(0,1){3}}
\put(26,49){\makebox(0,0) [l] {$\displaystyle{ \sim \sum_{n_{r}\lambda_{r},n_{s}\lambda_{s},k,k'}\frac{1}{\sqrt{[k,k']}} \; \Theta^{(2)}(
n_m\ell_m j_m, n_n\ell_n j_n, n_{m'}\ell_{m'} j_{m'}, n_{n'}\ell_{n'} j_{n'},n_{r}\ell_{r}j_r, n_{s}\ell_{s}j_{s}, \; \Xi)}$}}
\put(128,19){\vector(0,1){12}}
\put(128,30){\oval(5,5)[b]}
\put(128,19){\circle*{1.7}}
\put(123,27){\makebox(0,0)[t]{$ n_{m}j_{m}$}}
\put(138,19){\vector(0,1){12}}
\put(138,30){\oval(5,5)[b]}
\put(138,19){\circle*{1.7}}
\put(143,27){\makebox(0,0)[t]{$ n_{n}j_{n}$}}
\put(128,19.6){\line(10,0){10}}
\put(128,19.3){\line(10,0){10}}
\put(128,19){\line(10,0){10}}
\put(128,18.7){\line(10,0){10}}
\put(128,18.4){\line(10,0){10}}
\put(131,17){\line(0,4){4}}
\put(136,19.6){\vector(-1,0){3}}
\put(136,18.4){\vector(-1,0){3}}
\put(133,23){\makebox(0,0){$j_{12}$}}
\put(128,09){\line(0,1){10}}
\put(123,13){\makebox(0,0){$ n_{m'}j_{m'}$}}
\put(125,19){\makebox(0,0){$+$}}
\put(128,06){\vector(0,1){4}}
\put(128,08){\oval(5,5)[t]}
\put(138,09){\line(0,1){10}}
\put(143,13){\makebox(0,0){$ n_{n'}j_{n'}$}}
\put(141,19){\makebox(0,0){$-$}}
\put(138,06){\vector(0,1){4}}
\put(138,08){\oval(5,5)[t]}
\put(133,04){\makebox(0,0){$ A_{5}$}}
\put(72,27){\makebox(0,0)[t]{$j_m$}}
\put(98,27){\makebox(0,0)[t]{$j_{n}$}}
\put(78,25){\makebox(0,0)[t]{$j_r$}}
\put(92,25){\makebox(0,0)[t]{$j_s$}}
\put(80,29){\circle*{1.7}}
\put(78.5,31){\makebox(0,0){$+$}}
\put(90,29){\circle*{1.7}}
\put(92.5,31){\makebox(0,0){$-$}}
\put(80,29.6){\line(10,0){10}}
\put(80,29.3){\line(10,0){10}}
\put(80,29){\line(10,0){10}}
\put(80,28.7){\line(10,0){10}}
\put(80,28.4){\line(10,0){10}}
\put(83,27){\line(0,4){4}}
\put(88,29.6){\vector(-1,0){3}}
\put(88,28.4){\vector(-1,0){3}}
\put(85,32){\makebox(0,0){$k$}}
\put(70.5,19){\circle*{1.7}}
\put(99.5,19){\circle*{1.7}}
\put(70.5,19.6){\line(29,0){29}}
\put(70.5,19.3){\line(29,0){29}}
\put(70.5,19){\line(29,0){29}}
\put(70.5,18.7){\line(29,0){29}}
\put(70.5,18.4){\line(29,0){29}}
\put(83,17){\line(0,4){4}}
\put(88,19.6){\vector(-1,0){3}}
\put(88,18.4){\vector(-1,0){3}}
\put(85,23){\makebox(0,0){$j_{12}$}}
\put(80,09){\circle*{1.7}}
\put(78,07){\makebox(0,0){$+$}}
\put(90,09){\circle*{1.7}}
\put(92,07){\makebox(0,0){$-$}}
\put(80,09.6){\line(10,0){10}}
\put(80,09.3){\line(10,0){10}}
\put(80,09){\line(10,0){10}}
\put(80,08.7){\line(10,0){10}}
\put(80,08.4){\line(10,0){10}}
\put(83,07){\line(0,4){4}}
\put(88,09.6){\vector(-1,0){3}}
\put(88,08.4){\vector(-1,0){3}}
\put(85,12){\makebox(0,0){$k'$}}
\put(70,19){\line(1,1){10}}
\put(70,19){\line(1,-1){10}}
\put(100,19){\line(-1,1){10}}
\put(100,19){\line(-1,-1){10}}
\put(80,09){\line(0,20){20}}
\put(72,13){\makebox(0,0){$j_{m'}$}}
\put(80,14){\vector(0,1){3}}
\put(67.5,19){\makebox(0,0){$+$}}
\put(90,09){\line(0,20){20}}
\put(90,14){\vector(0,1){3}}
\put(98,13){\makebox(0,0){$j_{n'}$}}
\put(102.5,19){\makebox(0,0){$-$}}
\put(85,04){\makebox(0,0){$ A_{4}$}}
\put(37,17){\makebox(0,0) [l]{$\displaystyle{\times ~\sum_{j_{12}}}$}}
\end{picture}
\caption{One of the diagrams of the second-oder effective Hamiltonian.}
\label{d-b}
\end{center}
\end{figure}

The technique proposed in \cite{Meratal:86} leads to an effective Hamiltonian
$H_{eff}$ \cite{Meratal:85} in second order PT, which can be expressed as the sum
of terms corresponding to four classes of Feynman diagrams: vacuum, one-electron, two-electron and
three-electron. We illustrate the construction by considering two Feynman
 diagrams: $A_{1}$ (Fig.~\ref{d-a}) corresponds a first-order two-electron diagram and
$A_{3}$ (Fig.~\ref{d-b}) corresponds a second-order two-electron diagram. 
The following notation are used in these figures - 
$m \equiv n_m \ell_m j_{m} \,$, $\, m' \equiv n_{m'} \ell_{m'} j_{m'} \,$,
$n \equiv n_n \ell_n j_{n} \,$, $\, n' \equiv n_{n'} \ell_{n'} j_{n'} \,$, 
$r \equiv n_r \ell_r j_{r}  \,$, $s \equiv n_s \ell_s j_{s}  \,$, 
and $\, a \equiv n_a \ell_a j_{a}$.
The amplitude $\Theta ^{(I)}\left( 
n_m \ell_m j_{m}, n_n \ell_n j_{n} ,
n_{m'} \ell_{m'} j_{{m'}}, n_{n'} \ell_{n'} j_{n'}, \;
\Xi \right) $
is proportional to the two-electron one  or two submatrix elements (the effective interaction strength) of a 
two-particle physical operator~\cite{Gaigalas_1996,Gaigalas_1997,Gaigalas:89}
\begin{equation}
\label{eq:two-sterngth_1}
\Theta ^{(1)}\left( 
n_m \ell_m j_{m}, n_n \ell_n j_{n} ,
n_{m'} \ell_{m'} j_{m'}, n_{n'} \ell_{n'} j_{n'}, \;
\Xi \right) \; \sim \;
   \redmem{(n_m \ell_m)\, j_{m} \; (n_n \ell_n)\, j_{n} }{\, g_{12} \,}
	{(n_{m'} \ell_{m'})\, j_{m'} \; (n_{n'} \ell_{n'})\, j_{n'}}
\end{equation}
and
\begin{eqnarray}
\label{eq:two-sterngth_2}
\Theta ^{(2)}\left( 
n_m \ell_m j_{m}, n_n \ell_n j_{n} ,
n_{m'} \ell_{m'} j_{m'}, n_{n'} \ell_{n'} j_{n'}, 
n_{r} \ell_{r} j_{r}, n_{s} \ell_{s} j_{s}, \;
\Xi \right)  
	\nonumber \\
\sim \;
   \redmem{(n_m \ell_m)\, j_{m} \; (n_n \ell_n)\, j_{n} }{\, g_{12} \,}
	{(n_{r} \ell_{r})\, j_{r} \; (n_{s} \ell_{s})\, j_{s}} \;
	  \redmem{(n_r \ell_r)\, j_{r} \; (n_s \ell_s)\, j_{s} }{\, g_{12} \,}
	{(n_{m'} \ell_{m'})\, j_{m'} \; (n_{n'} \ell_{n'})\, j_{n'}},
\end{eqnarray}
where $\Xi$ is an array of intermediate coupling. 

With the first Feynman diagram (A1) {\sc Grasp} package is already dealing for methods described in sections~\ref{sec:GRASP_proc} and \ref{sec:ZFOm}.
In Fig.~\ref{d-a}, the
left hand side shows the interaction diagram; the dotted line corresponds to a
two-electron interaction. This is decomposed into the
product of a reduced matrix elements composed of radial integrals and
other algebraic expressions with a coupled tensor operator
\begin{equation}
\label{eq:ef-b}
A_{2} := \left[\left[a^{(j_{m})} \times \tilde{a}^{(j_{m'})}\right]^{(k)} \times \left[a^{(j_{n})} \times \tilde{a}^{(j_{n'})}\right]^{(k)}\right]^{(0)}
\end{equation}
defined by the diagram $A_{2}$ in the right side of Fig. ~\ref{d-a}. In Eq. (\ref{eq:ef-b})
$a^{(j)}$ is electron creation operator. Tensor $\tilde{a}^{(j)}$
is defined \cite{RudKan:84a,Rud:97a}
\begin{equation}
\label{eq:ef-c}
\tilde{a}^{(j)}_{m_{j}} = (-1)^{j-m_j}{a}^{\dagger \, (j)}_{-m_{j}},
\end{equation}
where ${a}^{\dagger \, (j)}_{-m_{j}}$ is electron annihilation operator.

The second Feynman diagram (A3) is coming from second order of perturbation theory and such type of diagrams (second order of effective hamiltonian) must be implemented in the {\sc Grasp}.
Similarly, the Feynman diagram $A_{3}$ is
expressed as the product of a more complicated second-order perturbation
expression with a recoupling coefficient $A_{4}$, and a coupled tensor
operator $A_{5}$ with the same structure as $A_{2}$ (see Fig.~\ref{d-a}).
So, the diagrams are different, but spin-angular part is the same: $A_2 = A_5$. Therefore  this formulation of perturbation theory allows us significantly simplify calculation
of effective operator with any number of open subshells for complex atoms and ions by using standard library of {\sc Grasp} for spin-angular integration~\cite{Gaigalas:2022}. 

\subsection{Relativistic second order effective Hamiltonian of an atom or an ion in irreducible tensorial form}
\label{sec:seconOrder}

Originally, Rayleigh-Schr\"odinger perturbation theory in irreducible tensorial form
was formulated in nonrelativistic atomic theory which is using $LS$-coupling~\cite{Meratal:86,Gaigalas:89}. Here we derive from the scratch such part of second order of effective operator
which correspond to core-valence correlations in relativistic atomic theory which is using $jj$-coupling. For this we were using the same technique as in ~\cite{Meratal:86,Gaigalas:89} that is 
a generalized graphical approach~\cite{Gaigatal:85} and the second quantization in coupled tensorial form \cite{RudKan:84a}. For this type of correlations we got two one- particle Feynman diagrams and 
four two-particle Feynman diagrams which contribution to CV correlations are significant. These diagrams are expressed via scalar, multielectron operators which operate only in the space of the wave functions of open subshells and are not explicitly a function of the magnetic quantum numbers. Such form of an effective operator is very convenient in practical calculation of the energy spectra of atoms and ions
with open electron subshells, because we can use all advantage of all set of methods described in~\cite{Gaigalas_1997}.
For the deriving expression of effective operator and applying it to calculation, the
definition of module spaces must be determined.
We are using the same definition as 
in~\cite{LindgrenBook:82,Meratal:86,Gaigalas:89}. The effective operator itself is acting
to the $P$-space, similar as in i) in subsection~\ref{sec:ZFOm}. The complementary part of the functional space is orthogonal space or the $Q$-space, similar as in ii) in subsection~\ref{sec:ZFOm}.

The module $P$-space in addition to the $Q$-space, are normally broken down into three sets -$F$, $F'$ and $G$. The $F$ set includes the orbitals, which describe the filled (core) subshells encountered in all CSFs. Meanwhile $F'$ set includes the orbitals, which corresponds to the unfilled (valence) subshells in some of these CSFs or at least in one of them. The $G$ set is built from the orbitals belonging to virtual subshells, which do not belong to $F$ or $F'$ and are used for generating the CSF for counting correlations effects. 
To distinguish between these spaces in Feynman diagrams and algebraic notations, we use the following notations:
\begin{eqnarray*}
\label{eq:space}
   \begin{array}{ll}
  \mbox{indexes  } \; a \; \mbox{  and  }\;  b             & \mbox{for  } \; F  \; \; \mbox{set},  \\
  \mbox{indexes  } \; n, \; n',\;  m,\;  \mbox{and }\;  m' & \mbox{for  } \; F' \;    \mbox{set}, \\
  \mbox{indexes  } \; r \; \mbox{  and  }\;  s             & \mbox{for  } \; G  \; \; \mbox{set}.  \\
   \end{array}
\end{eqnarray*}
In more detail about each contribution of the Feynman diagram to core-valence correlations we will discus in following subsections.  

\subsubsection{The first type of core-valence correlations}
\label{subsec:first_type}

\begin{figure*}
\begin{center}
\setlength{\unitlength}{1mm}
\begin{picture}(180,43)
\thicklines
\put(10,30){\line(0,1){10}}
\put(10,38){\vector(0,1){2}}
\put(10,40){\vector(0,1){2}}
\put(12.5,38){\makebox(0,0)[t]{\small{$m$}}}
\multiput(10,30)(1,0){10}{\circle*{0.35}}
\put(10,20){\line(0,10){10}}
\put(10,24){\vector(0,1){2}}
\put(8,25){\makebox(0,0)[t]{\small{$r$}}}
\qbezier(20,20)(15,25)(20,30)
\put(17.6,25){\vector(0,1){2}}
\put(15.5,25){\makebox(0,0)[t]{\small{$s$}}}
\qbezier(20,20)(25,25)(20,30)
\put(22.4,25){\vector(0,-1){2}}
\put(24.5,25){\makebox(0,0)[t]{\small{$a$}}}
\multiput(10,20)(1,0){10}{\circle*{0.35}}
\put(10,10){\line(0,1){10}}
\put(10,10){\vector(0,1){3}}
\put(12.5,13){\makebox(0,0){\small{$m^{\prime}$}}}
\put(10,08){\vector(0,1){3}}
\put(15,04){\makebox(0,0){$\text{CV}_{1}$}}
\put(27,25){\makebox(0,0) [l] {$\displaystyle{ = 
	- \sum_{m, m^{\prime}}~\frac{1}{\sqrt{\left[ j_{m'} \right]}}~\left[\;  a^{\left( j_m \right) }  \times
  \tilde a^{\left( j_{m^{\prime}} \right) } \; \right] ^{\left( 0 \right)}
  ~\sum_{r , s , a}~\frac{\left( -1 \right)^{j_{m}+j_r+j_s+j_a}}{\left( \varepsilon_{m'}+\varepsilon_a-\varepsilon_r-\varepsilon_s \right)}
}$}}
\put(37,10){\makebox(0,0) [l] {$\displaystyle{ \times
 \sum_{k,k'}~\frac{\delta \left( k, k' \right)}{\sqrt{\left[k,k'\right]}}
	~X_{k}(m a, r s) ~ X_{k'}(r s, m^{\prime} a)}$}} 

\end{picture}
\caption{The CV Feynman diagram of the second-order effective Hamiltonian for direct part of excitation 
$    (n_{a} \ell_{a})\, j_{a}^{2j_a+1} \, (n_{m'} \ell_{m'})\, j_{m'}^{w_m'} 
   \rightarrow (n_{a} \ell_{a})\, j_{a}^{2j_a} \; (n_{m'} \ell_{m'})\, j_{m'}^{w_m'-1} \; (n_{s} \ell_{s})\, j_{s} \; (n_{r} \ell_{r})\, j_{r}$.}
\label{CV_1}
\end{center}
\end{figure*}

Here we will discus about first type of core-valence correlations which are presented by one-particle Feynman diagrams in Fig.~\ref{CV_1} and Fig.~\ref{CV_2}
\begin{equation}
\label{eq:CV-a}
    (n_{a} \ell_{a})\, j_{a}^{2j_a+1} \, (n_{m'} \ell_{m'})\, j_{m'}^{w_m'} 
   \rightarrow (n_{a} \ell_{a})\, j_{a}^{2j_a} \; (n_{m'} \ell_{m'})\, j_{m'}^{w_m'-1} \; (n_{s} \ell_{s})\, j_{s} \; (n_{r} \ell_{r})\, j_{r} .
\end{equation}

\begin{figure*}
\begin{center}
\setlength{\unitlength}{1mm}
\begin{picture}(180,43)
\thicklines
\put(10,30){\line(0,1){10}}
\put(10,38){\vector(0,1){2}}
\put(10,40){\vector(0,1){2}}
\put(12.5,38){\makebox(0,0)[t]{\small{$m$}}}
\put(20,30){\line(-1,-1){10}}
\put(18,22){\vector(-1,1){2}}
\put(11,27){\makebox(0,0)[t]{\small{$r$}}}
\multiput(10,30)(1,0){10}{\circle*{0.35}}
\put(20,20){\line(0,10){10}}
\put(20,25){\vector(0,-1){2}}
\put(22.5,25){\makebox(0,0)[t]{\small{$a$}}}
\multiput(10,20)(1,0){10}{\circle*{0.35}}
\put(10,10){\line(0,1){10}}
\put(10,10){\vector(0,1){3}}
\put(12.5,13){\makebox(0,0){\small{$m^{\prime}$}}}
\put(10,08){\vector(0,1){3}}
\put(20,20){\line(-1,1){10}}
\put(16,26){\vector(1,1){2}}
\put(11,24){\makebox(0,0)[t]{\small{$s$}}}
\put(15,04){\makebox(0,0){$\text{CV}_{2}$}}
\put(27,25){\makebox(0,0) [l] {$\displaystyle{ = -
\sum_{m, m^{\prime}}~\frac{1}{\sqrt{\left[ j_{m'} \right]}}~\left[\;  a^{\left( j_m \right) } \times
  \tilde a^{\left( j_{m^{\prime}} \right) } \; \right] ^{\left( 0 \right)}
  ~\sum_{r , s , a}~\frac{\left( -1 \right)^{j_m+j_{r}+j_s+j_a}}{\left( \varepsilon_{m'}+\varepsilon_a-\varepsilon_r-\varepsilon_s \right)}
}$}}
\put(37,10){\makebox(0,0) [l] {$\displaystyle{ \times
  \sum_{k,k'}
  \left\{
    \begin{array}{ccc}
      k  & j_{m} & j_{r} \\
      k' & j_{a} & j_{s}
    \end{array} \right\} 
	~X_{k}(m a, r s) ~ X_{k'}(s r, m^{\prime} a)}$}}
\end{picture}
\caption{The CV Feynman diagram of the second-order effective Hamiltonian for exchange part of excitation 
$(n_{a} \ell_{a})\, j_{a}^{2j_a+1} \, (n_{m'} \ell_{m'})\, j_{m'}^{w_m'} 
   \rightarrow (n_{a} \ell_{a})\, j_{a}^{2j_a} \; (n_{m'} \ell_{m'})\, j_{m'}^{w_m'-1} \; (n_{s} \ell_{s})\, j_{s} \; (n_{r} \ell_{r})\, j_{r}$.}
\label{CV_2}
\end{center}
\end{figure*}

Each second order Feynman diagram's expression of perturbation theory has the energy denominator
$D = \sum \left( \varepsilon_{\text{down}} - \varepsilon_{\text{up}} \right)$,
where $\varepsilon_{\text{down}}$ ($\varepsilon_{\text{up}}$) is the single-particle eigenvalue associated with the down- (up-)
going orbital lines to (from) the lowest interaction line of diagram. For example the denominator for  CV$_1$ diagram in Fig.~\ref{CV_1} is
\begin{equation}
\label{eq:denominator}
D = \left( \varepsilon_{m'}+\varepsilon_a-\varepsilon_r-\varepsilon_s \right).
\end{equation}

Also the following notations are used in the expressions of these diagrams (see Fig.~\ref{CV_1} and Fig.~\ref{CV_2}):
\begin{equation}
\label{eq:deffX}
   X_{k}(i j, i' j') 
   = \redmem{\ell_i j_{i}}{\, C^{(k)} \,}{ \ell_{i'} j_{i'}}
     \redmem{\ell_j j_{j}}{\, C^{(k)} \,}{ \ell_{j'} j_{j'}} 
R^{k}(n_i j_i \, n_jj_j, \, n_{i'}j_{i'} \, n_{j'}j_{j'} ),
\end{equation}
where $R^{k}\left(n_i j_i \, n_jj_j, \, n_{i'}j_{i'} \, n_{j'}j_{j'} \right)$ is the radial integral 
of electrostatic interaction between electrons~\cite[(89) and (90)]{Fisetal:16a} and 
$\redmem{\ell_i j_{i}}{\, C^{(k)} \,}{ \ell_{i'} j_{i'}}$ is the reduced matrix element of the irreducible tensor operator $C^{(k)}$ in $jj$-coupling.

The spin-angular part of the diagrams CV$_1$ and CV$_2$ is the same, it has
normal order of operators accretion and annihilation, it is in the irreducible tensorial form, and itself is a scalar operator
\begin{equation}
\label{eq:s-a_CV1}
 \left[\;  a^{\left( j_m \right) }  \times
  \tilde a^{\left( j_{m^{\prime}} \right) } \; \right] ^{\left( 0 \right)},
\end{equation}
which acts only to $F'$ set of $P$-space. The spin-angular part of these Feynman diagrams is the same as for Dirac operator ${\cal H}_{D}$~\cite{grantBook:07,Fisetal:16a}, with reduced matrix element expressed through spin-angular coefficient $t_{ab}^{\alpha\beta}$ and radial integrals $I(a, \, b)$~\cite{Gaigalas:2022}. But interacting strength is more complicated having additional coefficient such as $6j$-symbol and additional summation over $G$ set of orbitals. For example, CV$_2$ diagram has
\begin{equation*}
\label{eq:inter_strength_CV1}
 \frac{1}{\sqrt{\left[ j_{m'} \right]}} \sum_{r , s , a}~\frac{\left( -1 \right)^{j_{r}+j_s-j_m-j_a}}{\left( \varepsilon_{m'}+\varepsilon_a-\varepsilon_r-\varepsilon_s \right)}
 \sum_{k,k'}
  \left\{
    \begin{array}{ccc}
      k  & j_{m} & j_{r} \\
      k' & j_{a} & j_{s}
    \end{array} \right\} 
	~X_{k}(m a, r s) ~ X_{k'}(s r, m^{\prime} a).
\end{equation*}

The same Feynman diagrams from Fig.~\ref{CV_1} and Fig.~\ref{CV_2} described the core-valence correlations in case $r \equiv s$ as well
\begin{equation}
\label{eq:CV-b}
    (n_{a} \ell_{a})\, j_{a}^{2j_a+1} \, (n_{m'} \ell_{m'})\, j_{m'}^{w_m'} 
   \rightarrow (n_{a} \ell_{a})\, j_{a}^{2j_a} \; (n_{m'} \ell_{m'})\, j_{m'}^{w_m'-1} \; (n_{s} \ell_{s})\, j_{s}^2 .
\end{equation}

\subsubsection{The second type of core-valence correlations}
\label{subsec:second_type}

The second type of core-valence correlations 
\begin{equation}
\label{eq:CV-c}
    (n_{a} \ell_{a})\, j_{a}^{2j_a+1} \, (n_{m'} \ell_{m'})\, j_{m'}^{w_m'} \, (n_{n'} \ell_{n'})\, j_{n'}^{w_n'} 
   \rightarrow (n_{a} \ell_{a})\, j_{a}^{2j_a} \; (n_{m'} \ell_{m'})\, j_{m'}^{w_m'-1} \; (n_{n'} \ell_{n'})\, j_{n'}^{w_n'+1}
	 \; (n_{r} \ell_{r})\, j_{r} 
\end{equation}
can be expressed via one-particle (Fig.~\ref{CV_1} and \ref{CV_2})
and two-particle  
(Fig.~ \ref{CV_3}, \ref{CV_4}, \ref{CV_5}, and \ref{CV_6}) Feynman diagrams.

\begin{figure*}
\begin{center}
\setlength{\unitlength}{1mm}
\begin{picture}(180,58)
\thicklines
\put(10,45){\line(0,1){10}}
\put(10,53){\vector(0,1){2}}
\put(10,55){\vector(0,1){2}}
\put(12.5,53){\makebox(0,0)[t]{\small{$m$}}}
\put(20,45){\line(1,-1){10}}
\put(28,43){\vector(1,1){2.4}}
\put(27,42){\vector(1,1){2}}
\put(27,46){\makebox(0,0)[t]{\small{$n$}}}
\multiput(10,45)(1,0){10}{\circle*{0.35}}
\put(10,35){\line(0,10){10}}
\put(10,39){\vector(0,1){2}}
\put(8,40){\makebox(0,0)[t]{\small{$r$}}}
\put(20,35){\line(0,10){10}}
\put(20,40){\vector(0,-1){2}}
\put(18,40){\makebox(0,0)[t]{\small{$a$}}}
\multiput(10,35)(1,0){10}{\circle*{0.35}}
\put(10,25){\line(0,1){10}}
\put(10,25){\vector(0,1){3}}
\put(12.5,28){\makebox(0,0){\small{$m^{\prime}$}}}
\put(10,23){\vector(0,1){3}}
\put(20,35){\line(1,1){10}}
\put(29,36){\vector(-1,1){2.4}}
\put(26.2,35.4){\makebox(0,0){\small{$n'$}}}
\put(30,35){\vector(-1,1){2}}
\put(17,19){\makebox(0,0){$\text{CV}_{3}$}}
\put(33,40){\makebox(0,0) [l] {$\displaystyle{ = 
	\sum_{m, m^{\prime}}~\sum_{n, n^{\prime}}~\sum_{j_{12}}~~\sqrt{\left[ j_{12} \right]}~ 
		\left[\left[\;  a^{\left( j_m \right) }  \times 
  \tilde a^{\left( j_{m^{\prime}} \right) } \; \right] ^{\left( j_{12} \right)} \times 
	\left[\; \tilde a^{\left( j_{n^{\prime}} \right) }  \times 
  a^{\left( j_{n} \right) } \; \right] ^{\left( j_{12} \right)} \right]^{\left( 0 \right)}	
	}$}}
\put(43,25){\makebox(0,0) [l] {$\displaystyle{ \times
\sum_{r, a}~\sum_{k,k'}~\frac{\left( -1 \right)^{j_{m}+j_{m^{\prime}}+k+k^{\prime}}}
{\left( \varepsilon_{m'}+\varepsilon_a-\varepsilon_r-\varepsilon_n \right)}
	\left\{
    \begin{array}{ccc}
      k  & k' & j_{12} \\
      j_{m'} & j_{m} & j_{r}
     \end{array}
	\right\} 
  \left\{
    \begin{array}{ccc}
      k  & k' & j_{12} \\
      j_{n} & j_{n'} & j_{a}
    \end{array}
	\right\}
	}$}}
\put(43,10){\makebox(0,0) [l] {$\times \; X_{k}(m a, r n') ~ X_{k'}(r n, m^{\prime} a)$}}
\end{picture}
\caption{The CV Feynman diagram of the second-order effective Hamiltonian for direct part of the second type of core-valence correlations 
$(n_{a} \ell_{a})\, j_{a}^{2j_a+1} \, (n_{m'} \ell_{m'})\, j_{m'}^{w_m'} \, (n_{n'} \ell_{n'})\, j_{n'}^{w_n'} 
 \rightarrow (n_{a} \ell_{a})\, j_{a}^{2j_a} \; (n_{m'} \ell_{m'})\, j_{m'}^{w_m'-1} \; (n_{n'} \ell_{n'})\, j_{n'}^{w_n'+1} \; (n_{r} \ell_{r})\, j_{r}$.}
\label{CV_3}
\end{center}
\end{figure*}

\begin{figure*}
\begin{center}
\setlength{\unitlength}{1mm}
\begin{picture}(180,43)
\thicklines
\put(10,30){\line(0,1){10}}
\put(10,38){\vector(0,1){2}}
\put(10,40){\vector(0,1){2}}
\put(7.5,38){\makebox(0,0)[t]{\small{$m$}}}
\put(20,30){\line(0.5,-1){5}}
\put(17.5,38){\makebox(0,0)[t]{\small{$n$}}}
\multiput(10,30)(1,0){10}{\circle*{0.35}}
\put(10,20){\line(0,10){10}}
\put(25,25){\makebox(0,0)[t]{\small{$r$}}}
\put(15,38){\vector(0,1){2}}
\put(15,40){\vector(0,1){2}}
\put(15,20){\line(0,10){20}}
\put(18.5,27){\vector(-0.5,-1){2}}
\put(19.5,25){\makebox(0,0)[t]{\small{$a$}}}
\multiput(15,20)(1,0){10}{\circle*{0.35}}
\put(10,10){\line(0,1){10}}
\put(10,10){\vector(0,1){3}}
\put(7.5,13){\makebox(0,0){\small{$m^{\prime}$}}}
\put(10,08){\vector(0,1){3}}
\put(15,20){\line(0.5,1){5}}
\put(25,10){\line(0,1){10}}
\put(25,10){\vector(0,1){3}}
\put(25,08){\vector(0,1){3}}
\put(22.5,13){\makebox(0,0){\small{$n^{\prime}$}}}
\put(24,22){\vector(-0.5,1){2}}
\put(15,04){\makebox(0,0){$\text{CV}_{4}$}}
\put(30,25){\makebox(0,0) [l] {$\displaystyle{ = - 
	\sum_{m, m^{\prime}}~\sum_{n, n^{\prime}}~\sum_{k}~~\frac{1}{\sqrt{\left[ k \right]}}~ 
		\left[\left[\;  a^{\left( j_m \right) }  \times 
  \tilde a^{\left( j_{m^{\prime}} \right) } \; \right] ^{\left( k \right)} \times 
	\left[\; a^{\left( j_{n} \right) }  \times 
  \tilde a^{\left( j_{n^{\prime}} \right) } \; \right] ^{\left( k \right)} \right]^{\left( 0 \right)}	
	}$}}
\put(40,10){\makebox(0,0) [l] {$\displaystyle{ \times
\sum_{r, a}~\sum_{k'}~\frac{\left( -1 \right)^{j_{a}+j_{r}+k}}
{\left( \varepsilon_{n'}+\varepsilon_a-\varepsilon_r-\varepsilon_n \right)}
	\left\{
    \begin{array}{ccc}
      k  & j_{n} & j_{n'} \\
      k' & j_{r} & j_{a}  
     \end{array}
	\right\}  
	~X_{k}(m a, m' r) ~ X_{k'}(n r, a n^{\prime})}$}}
\end{picture}
\caption{The CV Feynman diagram of the second-order effective Hamiltonian for exchange part of the second type of core-valence correlations 
$(n_{a} \ell_{a})\, j_{a}^{2j_a+1} \, (n_{m'} \ell_{m'})\, j_{m'}^{w_m'} \, (n_{n'} \ell_{n'})\, j_{n'}^{w_n'} 
   \rightarrow (n_{a} \ell_{a})\, j_{a}^{2j_a} \; (n_{m'} \ell_{m'})\, j_{m'}^{w_m'-1} \; (n_{n'} \ell_{n'})\, j_{n'}^{w_n'+1}
	 \; (n_{r} \ell_{r})\, j_{r}$.}
\label{CV_4}
\end{center}
\end{figure*}

\begin{figure*}
\begin{center}
\setlength{\unitlength}{1mm}
\begin{picture}(180,43)
\thicklines
\put(10,30){\line(0,1){10}}
\put(10,38){\vector(0,1){2}}
\put(10,40){\vector(0,1){2}}
\put(7.5,38){\makebox(0,0)[t]{\small{$m$}}}
\put(10,30){\line(0.5,-1){5}}
\put(28,38){\makebox(0,0)[t]{\small{$n$}}}
\multiput(10,30)(1,0){10}{\circle*{0.35}}
\put(20,20){\line(0,10){10}}
\put(10,26){\makebox(0,0)[t]{\small{$r$}}}
\put(25,38){\vector(0,1){2}}
\put(25,40){\vector(0,1){2}}
\put(25,20){\line(0,10){20}}
\put(18.5,27){\vector(-0.5,-1){2}}
\put(15.5,26){\makebox(0,0)[t]{\small{$a$}}}
\multiput(15,20)(1,0){10}{\circle*{0.35}}
\put(20,10){\line(0,1){10}}
\put(20,10){\vector(0,1){3}}
\put(17.5,13){\makebox(0,0){\small{$m^{\prime}$}}}
\put(20,08){\vector(0,1){3}}
\put(15,20){\line(0.5,1){5}}
\put(25,10){\line(0,1){10}}
\put(25,10){\vector(0,1){3}}
\put(25,08){\vector(0,1){3}}
\put(27.5,13){\makebox(0,0){\small{$n^{\prime}$}}}
\put(14,22){\vector(-0.5,1){2}}
\put(15,04){\makebox(0,0){$\text{CV}_{5}$}}
\put(30,25){\makebox(0,0) [l] {$\displaystyle{ = - 
	\sum_{m, m^{\prime}}~\sum_{n, n^{\prime}}~\sum_{k'}~~\frac{1}{\sqrt{\left[ k' \right]}}~ 
		\left[\left[\;  a^{\left( j_m \right) }  \times 
  \tilde a^{\left( j_{m^{\prime}} \right) } \; \right] ^{\left( k' \right)} \times 
	\left[\; a^{\left( j_{n} \right) }  \times 
  \tilde a^{\left( j_{n^{\prime}} \right) } \; \right] ^{\left( k' \right)} \right]^{\left( 0 \right)}	
	}$}}
\put(40,10){\makebox(0,0) [l] {$\displaystyle{ \times
\sum_{r, a}~\sum_{k}~\frac{\left( -1 \right)^{j_{a}+j_{r}+k^{\prime}}}
{\left( \varepsilon_{n'}+\varepsilon_a-\varepsilon_r-\varepsilon_n \right)}
	\left\{
    \begin{array}{ccc}
      k'  & j_{m} & j_{m'} \\
      k & j_{a} & j_{r}  
     \end{array}
	\right\}  
	~X_{k}(m a, r m') ~ X_{k'}(r n, a n^{\prime})}$}}
	\end{picture}
\caption{The CV Feynman diagram of the second-order effective Hamiltonian for exchange part of the second type of core-valence correlations 
$(n_{a} \ell_{a})\, j_{a}^{2j_a+1} \, (n_{m'} \ell_{m'})\, j_{m'}^{w_m'} \, (n_{n'} \ell_{n'})\, j_{n'}^{w_n'} 
   \rightarrow (n_{a} \ell_{a})\, j_{a}^{2j_a} \; (n_{m'} \ell_{m'})\, j_{m'}^{w_m'-1} \; (n_{n'} \ell_{n'})\, j_{n'}^{w_n'+1}
	 \; (n_{r} \ell_{r})\, j_{r}$.}
\label{CV_5}
\end{center}
\end{figure*}

\begin{figure*}
\begin{center}
\setlength{\unitlength}{1mm}
\begin{picture}(180,43)
\thicklines
\put(10,30){\line(0,1){10}}
\put(10,38){\vector(0,1){2}}
\put(10,40){\vector(0,1){2}}
\put(12.5,38){\makebox(0,0)[t]{\small{$m$}}}
\multiput(10,30)(1,0){10}{\circle*{0.35}}
\put(10,10){\line(0,1){10}}
\put(10,10){\vector(0,1){3}}
\put(12.5,13){\makebox(0,0){\small{$m^{\prime}$}}}
\put(10,08){\vector(0,1){3}}
\put(10,20){\line(0,10){10}}
\qbezier(20,20)(15,25)(20,30)
\put(17.6,25){\vector(0,1){2}}
\put(15.5,25){\makebox(0,0)[t]{\small{$r$}}}
\qbezier(20,20)(25,25)(20,30)
\put(22.4,25){\vector(0,-1){2}}
\put(24.5,25){\makebox(0,0)[t]{\small{$a$}}}
\multiput(20,20)(1,0){10}{\circle*{0.35}}
\put(30,30){\line(0,1){10}}
\put(30,38){\vector(0,1){2}}
\put(30,40){\vector(0,1){2}}
\put(32.5,38){\makebox(0,0)[t]{\small{$n$}}}
\put(30,20){\line(0,10){10}}
\put(30,08){\vector(0,1){3}}
\put(30,10){\line(0,1){10}}
\put(30,10){\vector(0,1){3}}
\put(32.5,13){\makebox(0,0){\small{$n^{\prime}$}}}
\put(15,04){\makebox(0,0){$\text{CV}_{6}$}}
\put(35,25){\makebox(0,0) [l] {$\displaystyle{ = 
	- \sum_{m, m^{\prime}}~\sum_{n, n^{\prime}}~\sum_{k,k'}~~\frac{\delta \left(k,k'\right)}{\left[ k \right]\sqrt{\left[ k \right]}}~
	\left[\left[\;  a^{\left( j_m \right) }  \times 
  \tilde a^{\left( j_{m^{\prime}} \right) } \; \right] ^{\left( k \right)} \times 
	\left[\;  a^{\left( j_n \right) }  \times 
  \tilde a^{\left( j_{n^{\prime}} \right) } \; \right] ^{\left( k \right)} \right]^{\left( 0 \right)}
	}$}}
\put(45,10){\makebox(0,0) [l] {$\displaystyle{ \times
\sum_{r, a}~\frac{\left( -1 \right)^{j_{a}+j_{r}+k}}
{\left( \varepsilon_{n'}+\varepsilon_a-\varepsilon_r-\varepsilon_n \right)}
	~X_{k}(m a, m' r) ~ X_{k'}(r n, a n^{\prime})}$}}
\end{picture}
\caption{The CV Feynman diagram of the second-order effective Hamiltonian for exchange part of the second type of core-valence correlations 
$(n_{a} \ell_{a})\, j_{a}^{2j_a+1} \, (n_{m'} \ell_{m'})\, j_{m'}^{w_m'} \, (n_{n'} \ell_{n'})\, j_{n'}^{w_n'} 
   \rightarrow (n_{a} \ell_{a})\, j_{a}^{2j_a} \; (n_{m'} \ell_{m'})\, j_{m'}^{w_m'-1} \; (n_{n'} \ell_{n'})\, j_{n'}^{w_n'+1}
	 \; (n_{r} \ell_{r})\, j_{r}$.}
\label{CV_6}
\end{center}
\end{figure*}

It need empathize that the Feynman diagrams CV$_3$, CV$_4$, CV$_5$, and CV$_6$ are not in a normal order of accretion and annihilation operators. These operators order in tensorial product is the same as Eq. (\ref{eq:ef-b})
for CV$_4$, CV$_5$, and CV$_6$ meantime the CV$_3$ is expressed via 
\begin{equation}
\label{eq:s-a_CV3}
\left[ \left[ a^{\left( j_m \right)}  \times \tilde{a}^{\left( j_{m'} \right)}  \right]^{\left( j_{12} \right)} \times 
\left[ \tilde{a}^{\left( j_{n'} \right)}  \times a^{\left( j_n \right)} \right]^{\left( j_{12} \right)} \right]^{\left( 0 \right)}.
\end{equation}
All these operators in Eq. (\ref{eq:ef-b}) and Eq. (\ref{eq:s-a_CV3}) act only to $F'$ set of $P$-space. 

The spin-angular part of these Feynman diagrams is very similar as for 
Coulomb operator ~\cite{grantBook:07,Fisetal:16a}, 
which has been expressed in normal order of second quantization operators as (see Eq. (20)~\cite{Gaigalas_1996})
\begin{equation}
\label{eq:s-a_B3}
\left[ \left[ a^{\left( j_m \right)} \times a^{\left( j_{n} \right)}  \right]^{\left( k \right)} \times 
\left[ \tilde{a}^{\left( j_{m'} \right)}  \times \tilde{a}^{\left( j_{n'} \right)}  \right]^{\left( k \right)} \right]^{\left( 0 \right)}
\end{equation}
or in the combination of two parts in which second quantization operators are ordered as pairs of creation-annihilation operators as Eq. (\ref{eq:ef-b}) and Eq. (\ref{eq:s-a_CV1})
(see Eq. (21)~\cite{Gaigalas_1996}).

Therefore, program library~\cite{Gaigalas:2022} from {\sc Grasp} supports calculation of the spin-angular part of
CV$_4$, CV$_5$, and CV$_6$ Feynman diagrams.
In addition, it is very easy to redefine expression  (Eq. (\ref{eq:s-a_CV3}) to Eq. (\ref{eq:s-a_B3}) and Eq. (\ref{eq:s-a_CV1})) of CV$_3$ diagram for access calculation of its spin-angular part with program library~\cite{Gaigalas:2022}, too.

\subsubsection{The contribution of core-valence correlations to off-diagonal matrix elements}
\label{subsec:offdiagonal_type}

The main contribution of core-valence correlations to off-diagonal matrix elements is in the matrix element
$\redmem{(n_{m} \ell_{m})\, j_{m}^{w_m} \, (n_{n} \ell_{n})\, j_{n}^{w_n}}{\, \widehat{{\cal H}}^{(2)}_{Effective} \,}{(n_{m} \ell_{m})\, j_{m}^{w_m-2} \, (n_{n} \ell_{n})\, j_{n}^{w_n+2} }$.
The above contribution is derived from the excitation:
\begin{equation}
\label{eq:CV-off_Diagonal} 
(n_{a} \ell_{a})\, j_{a}^{2j_a+1} \, (n_{m'} \ell_{m'})\, j_{m'}^{w_m'} \, (n_{n'} \ell_{n'})\, j_{n'}^{w_n'}
   \rightarrow (n_{a} \ell_{a})\, j_{a}^{2j_a} \; (n_{m'} \ell_{m'})\, j_{m'}^{w_m'-1} \; (n_{n'} \ell_{n'})\, j_{n'}^{w_n'+1}
	 \; (n_{r} \ell_{r})\, j_{r}
\end{equation}
and can be described by the same two-particle Feynman diagrams Fig.~\ref{CV_3}, \ref{CV_4}, \ref{CV_5}, and \ref{CV_6} as before.

\section{Implementation of Rayleigh-Schr\"odinger perturbation theory in irreducible tensorial form in the {\sc Grasp}2018~\cite{grasp2018}}
\label{sec:PT_implementation}

The expressions of CV correlations
in second order of perturbation theory 
 presented in section~\ref{sec:seconOrder}
are exact and come directly from the theory. 
For implementation of this theory in RCI approach, we are making the following assumptions:
\begin{enumerate}

\item Infinity summations over $G$ space are replaced by limited summations if our ASF in zero order of perturbation theory is already sufficiently accurate. This is usually
true after RMCDHF calculation with the inclusion of VV correlations or sometimes additionally CV excitation from one extreme trunk orbit. The correctness of this assumption is also confirmed by the accurate results obtained using the Zero-first-order method in the {\sc Grasp} package~\cite{Gusatal:2017,Jonsson:17}, since this method is based on the Brillouin-Wigner perturbation theory, which is also subject to infinite summation, as is the case for the Rayleigh-Schr\"odinger perturbation theory.

\item In reality, the denominator of energy $D$ (see Eq. (\ref{eq:denominator})) in CV$_1$, CV$_2$, CV$_3$, CV$_4$,
CV$_5$, and CV$_6$ contains only one-electron energies. But if in the combination of RCI and perturbation theory the one electronic orbitals are not spectroscopic (e.g. correlation orbitals \cite{Fisetal:16a,Jonsson:23}), then the one-electron energies do not correspond to the values of the real one-electron energies and cannot be used in Eq. (\ref{eq:denominator}). Therefore
in this case, we consider that the one-electron energies in the denominator 
$D$ in CV$_1$, CV$_2$, CV$_3$, CV$_4$,
CV$_5$, and CV$_6$
are changed to the averaged energies 
$\overline{D} = \overline{E}\left(K^{'}\right)-\overline{E}\left(K\right)$.
$\overline{E}\left(K\right)$ is the averaged energy of a 
state 
for which calculations are performed. 
$\overline{E}\left(K^{'}\right)$ is average energy for admixed configuration $K'$.
This allows us to move energy denominator in front of summations over the orbitals belonging to
$F$ and $G$ set in expressions of second order of effective Hamiltonian~\cite{Bogetal:1997,Bogetal:1998}. This allows simplify the implementation theory in the {\sc Grasp} program package. It need to point out that $D$ and $\overline{D}$ are defined differently in papers~\cite{Meratal:86,Gaigalas:89} and~\cite{LindgrenBook:82,Bogetal:1997,Bogetal:1998} respectively. 
In the first case, the energy of admixed configuration $K'$ 
is with minus sign in the energy difference $D$, while in the second case $\overline{D}$, on the contrary, it has 
the energy of the state under consideration $K$.
We will also abide in the paper by these notations. Therefore, $D$ and $\overline{D}$ will be defined with opposite signs.

For calculation average energies $\overline{E}\left(K\right)$ and $\overline{E}\left(K^{'}\right)$ we can use the analytical expression~\cite[(42)]{Gaigalas:2022}
\begin{eqnarray}
\label{eq:simple_average}
\hspace{-2.0cm}
   \overline{E}
	=  \sum_{a} w_{a} \; I(a, \; a)
	+  \sum_{a} \frac{w_a\left( w_a - 1 \right)}{2}
	\left[ F^0 (a, \; a) 
	-\frac{2j_a+1}{2j_a} \sum_{k>0}
	  \left( \ell_{a} \ell_{a} k\right)
   \left(
   \begin{array}{ccc}
      j_a          & k & j_a \\
      -\frac{1}{2} & 0 & \frac{1}{2}
   \end{array}
   \right)^2
	 F^k (a, \; a)
	\right]
   \nonumber  \\[1ex]
\hspace{-1.5cm}	
   + \sum_{a<b} w_{a}w_{b}
	\left[  F^0 (a, \; b)
	- \sum_{k}
	 \left( \ell_{a} \ell_{b} k\right)
   \left(
   \begin{array}{ccc}
      j_a          & k & j_b \\
      -\frac{1}{2} & 0 & \frac{1}{2}
   \end{array}
   \right)^2
	G^k (a, \; b)
	\right],
\end{eqnarray}
where symbol $\left( \ell_{a} \ell_{b} k\right)$ means that quantities $\ell_{a}$, $\ell_{b}$, and $k$ obey the triangular condition with even perimeter, which usually is included in the definition of radial integral~\cite{Rud:97a,Kar:96a,Gaigalas:2022}.
\end{enumerate}
A discussion on the correctness of these two assumptions is described in section \ref{sec:PT_1_Test_Case}.

Brillouin-Wigner perturbation theory (with these assumptions) 
is already implemented in the {\sc Grasp} package~\cite{grasp2013,grasp2018} years ago.
Accurate atomic data (with spectroscopic accuracy) \cite{Gusatal:2017,Gaigalasetal:2020,Gaigalasetal:2021}
were obtained by using this method.
Since such combination of the Brillouin-Wigner perturbation theory with RCI method allows to include correlations effects 
with high accuracy, therefore
Rayleigh-Schr\"odinger perturbation theory (also other analogous perturbation theory option) combined with RCI and with
the same 
assumption should give accurate results, too. 
This can be confirmed by the investigation of energy spectra and transition data for different ionization degree of tungsten 
atom~\cite{Bogdanovich&Kisielius2012,Bogdanovich&Kisielius2013,Karpuskiene&Kisielius2022} or various other
ions~\cite{Kisielius_etal2015,Bog:2005,Bogetal:1999} including investigation of some metastable-level
along several isoelectronic sequences for the ions from $Z$ = 50 to $Z$ = 92~\cite{Karpuskiene_etal2013}
and additionally calculating radiative lifetimes~\cite{Aggarwal_etal2016,Karpetal:2004,Bogetal:2001}
by using similar approach~\cite{Bogetal:1997,Bogetal:1998} in nonrelativistic atomic theory.
Therefore all these two assumptions should not impact negatively for accuracy of CV correlations in calculation atomic properties with the combination
 CI approach with PT.
But it should be noted that it is
important that in zero order space the most important correlations would be included in this 
Rayleigh-Schr\"odinger perturbation theory~\cite{Meratal:86,Gaigalas:89} and RCI combination, as
it is done using RCI combined with Brillouin-Wigner perturbation theory.

It is more convenient to reformulated the formalism of perturbation theory presented in section~{\ref{sec:PT_Mano}} in such way that this allow us to use the spin-angular library~{\cite{Gaigalas:2022} without any modification and 
to make most easy implementation of this theory in the {\sc Grasp}2018 package~\cite{grasp2018} in general. 
This reformulation is presented in this section.

According to \cite{Bogetal:1997} the contribution of the admixed configurations from CV correlations can be added to usual energy of the 
term $\chi J$ of the configuration $K$ (the set of
$K \chi J$ and/or $K' \chi' J'$ are the CSFs in the relativistic approach)
 and can be
expressed as the energy $E_0 \left(K J\right)$, which does not depend on term, and the sum of product of Slater integrals and spin-angular coefficients, describing the interaction within open subshell and between them:
\begin{eqnarray}
\label{eq:BogEnergy}
\hspace*{-2.5cm}
   E\left(K \chi J \right)
	\nonumber \\
& &
   = E_0 \left(K J\right) + \Delta \mathcal{E}_0 \left(K J \right) 
	\nonumber \\  [0.2cm]
& &
	+ \; \sum_{n \ell j} \sum_{k>0} \widetilde{f}_k \left( \ell j^{w}, \; K \chi J  \right)
	\left[ \mathcal{F}^{k} \left( n \ell j, \; n \ell j \right)  
	+ \Delta \mathcal{F}^{k} \left( n \ell j, \; n \ell j \right) \right]
	\nonumber \\
& &
	+ \; \sum_{n \ell j} \sum_{n' \ell' j' > n \ell j} \left\{ \sum_{k>0} \widetilde{f}_k \left( \ell j^{w} \; \ell' j'^{w'},
	\; K \chi J  \right) \right.
\left[ \mathcal{F}^{k} \left( n \ell j, \; n '\ell' j' \right)  
	+ \Delta \mathcal{F}^{k} \left( n \ell j, \; n' \ell' j' \right) \right]
	\nonumber \\ [0.2cm]
& & 
	+ \sum_{k} \widetilde{g}_k \left( \ell j^{w} \; \ell' j'^{w'}, \; K \chi J  \right)	
\left[ \mathcal{G}^{k} \left( n \ell j, \; n '\ell' j' \right)  
	+ \Delta \mathcal{G}^{k} \left( n \ell j, \; n' \ell' j' \right) \right] 
	\nonumber \\ [0.2cm]
& &
	+ \sum_{k} \widetilde{v}_k \left( \ell j^{w} \; \ell' j'^{w'}, \ell j^{w-2} \; \ell' j'^{w'+2},
	\; K \chi J \; K' \chi' J \right)	
	\nonumber \\
& &
  \times 
\left. 
\left[ \mathcal{R}^{k} \left( n \ell j n \ell j, \; n '\ell' j' n '\ell' j' \right)  
	+ \Delta \mathcal{R}^{k} \left( n \ell j n \ell j, \; n' \ell' j' n '\ell' j' \right) \right] \right\} ,
\end{eqnarray}
where $\widetilde{f}_k$, $\widetilde{g}_k$, and $\widetilde{v}_k$ are spin-angular coefficients in which submatrix elements $\redmem{\ell j}{\, C^{(k)} \,}{ \ell^{\prime} j^{\prime}}$ are extracted. 
Therefore summation over $k$ runs over all 
possible values instead the values which satisfy the triangular condition $\left( \ell \ell^{\prime} k\right)$ as it is in the ordinary case Eq. (\ref{eq:simple_average}). The $\mathcal{F}^{k} \left( n \ell j, \; n '\ell' j' \right)$,
$\mathcal{G}^{k} \left( n \ell j, \; n '\ell' j' \right)$, and $\mathcal{R}^{k} \left( n \ell j n \ell j, \; n '\ell' j' n '\ell' j' \right)$ are generalized integrals of electrostatic interaction between electrons. The definition of $\mathcal{R}^{k} \left( n \ell j n \ell j, \; n '\ell' j' n '\ell' j' \right)$ is following
\begin{eqnarray}
\label{eq:BogRk}
\hspace*{-2.5cm}
   \mathcal{R}^{k}\left(i j, i' j'\right)
	\nonumber \\
& &
   = \left\{ \left[ 1 + \delta \left( i, j \right) \right]  \left[ 1 + \delta \left( i', j' \right) \right]  \right\} ^{-1/2}
	 \, R^{k}\left(n_i j_i \, n_jj_j, \, n_{i'}j_{i'} \, n_{j'}j_{j'} \right)
\nonumber \\
& &
	   \times \redmem{\ell_i j_{i}}{\, C^{(k)} \,}{ \ell_{i'} j_{i'}}
     \redmem{\ell_j j_{j}}{\, C^{(k)} \,}{ \ell_{j'} j_{j'}}, 
\end{eqnarray}
where $R^{k}\left(n_i j_i \, n_jj_j, \, n_{i'}j_{i'} \, n_{j'}j_{j'} \right)$ is the same radial integral as in 
Eq. (\ref{eq:deffX}). Definitions  $\mathcal{F}^{k} \left( n \ell j, \; n '\ell' j' \right)$,
$\mathcal{G}^{k} \left( n \ell j, \; n '\ell' j' \right)$ straight forward follows from Eq. (\ref{eq:BogRk}).

The contribution coming from the CV correlations of the configurations $K'$ to $E (K \chi J)$ in second order of perturbation theory
can be written from Eq. (\ref{eq:BogEnergy}) as
\begin{eqnarray}
\label{eq:BogEnergy_PT}
\hspace*{-2.5cm}
  \Delta E_{PT}
	\nonumber \\
& &
   =  \Delta \mathcal{E}_0 \left(K J \right) 
	\nonumber \\  [0.2cm]
& &
	+ \; \sum_{n \ell j} \sum_{k>0} \widetilde{f}_k \left( \ell j^{w}, \; K \chi J  \right)
	\Delta \mathcal{F}^{k} \left( n \ell j, \; n \ell j \right)
	\nonumber \\
& &
	+ \; \sum_{n \ell j} \sum_{n' \ell' j' > n \ell j} \left\{ \sum_{k>0} \widetilde{f}_k \left( \ell j^{w} \; \ell' j'^{w'},
	\; K \chi J  \right) \right.
 \Delta \mathcal{F}^{k} \left( n \ell j, \; n' \ell' j' \right) 
	\nonumber \\ [0.2cm]
& & 
	+ \sum_{k} \widetilde{g}_k \left( \ell j^{w} \; \ell' j'^{w'}, \; K \chi J  \right)	
 \Delta \mathcal{G}^{k} \left( n \ell j, \; n' \ell' j' \right)
	\nonumber \\ [0.2cm]
& &
\left.
	+ \sum_{k} \widetilde{v}_k \left( \ell j^{w} \; \ell' j'^{w'}, \ell j^{w-2} \; \ell' j'^{w'+2},
	\; K \chi J \; K' \chi' J \right)	
\Delta \mathcal{R}^{k} \left( n \ell j n \ell j, \; n' \ell' j' n '\ell' j' \right) \right\} .
\end{eqnarray}

\begin{table*}
\begin{center}
\begin{tabular}{|l|} \hline
$\Delta \mathcal{E}_0$ corrections \\ \hline \hline
\\
$\overbrace{(n_{a} \ell_{a})\, j_{a}^{2j_a+1}}^{\text{core subshells}} \, \overbrace{(n_{m} \ell_{m})\, j_{m}^{w_m}}^{\text{valence subshells}} \;
\rightarrow \; \overbrace{(n_{a} \ell_{a})\, j_{a}^{2j_a}}^{\text{core subshells}} \; \overbrace{(n_{m} \ell_{m})\, j_{m}^{w_m-1}}^{\text{valence subshells}} \; \overbrace{(n_{s} \ell_{s})\, j_{s} \; (n_{r} \ell_{r})\, j_{r}}^{\text{virtual subshells}}$ \\
\\
$-\frac{w_m}{\left[ j_m\right]} \, \left\{ \underbrace{ 2 (-1)^{j_r+j_s+j_m+j_a}\sum_{k} \,
  \mathcal{C}\left(k, \, m a, \, r s \right)}_{\text{from $\text{CV}_2$ Feynman diagram}}
	+ \underbrace{\sqrt{\left[ j_m\right]} \left[ \sqrt{\left[ j_r\right]} \; \mathcal{A}\left( 0, \, m a, \, s r \right)
	+ \sqrt{\left[ j_s\right]} \; \mathcal{A}\left( 0, \, m a, \, r s \right) \right]}_{\text{from $\text{CV}_1$ Feynman diagram}} \right\}$ \\
\\ \hline
\\
$ \overbrace{(n_{a} \ell_{a})\, j_{a}^{2j_a+1}}^{\text{core subshells}} \, \overbrace{(n_{m} \ell_{m})\, j_{m}^{w_m}}^{\text{valence subshells}} \;
\rightarrow  \; \overbrace{(n_{a} \ell_{a})\, j_{a}^{2j_a}}^{\text{core subshells}} \; \overbrace{(n_{m} \ell_{m})\, j_{m}^{w_m-1}}^{\text{valence subshells}} \; 
\overbrace{(n_{s} \ell_{s})\, j_{s}^{2}}^{\text{virtual subshells}}$ \\
\\
$\frac{2 w_m}{\left[ j_m \right]} \left\{ \underbrace{ (-1)^{j_m+j_a} \sum_{k} \mathcal{C}\left(k, \, ma, \, ss \right)}_{\text{from $\text{CV}_2$ Feynman diagram}}
\underbrace{-\sqrt{\left[ j_m, j_s \right]} \; \mathcal{A}\left( 0, \, ma, \, ss \right)}_{\text{from $\text{CV}_1$ Feynman diagram}} \right\}$  \\ 
\\ \hline \hline
 \\
$\overbrace{(n_{a} \ell_{a})\, j_{a}^{2j_a+1}}^{\text{core subshells}} \, \overbrace{(n_{m} \ell_{m})\, j_{m}^{w_m} \; (n_{n} \ell_{n})\, j_{n}^{w_n}}^{\text{valence subshells}} \;
\rightarrow \; \overbrace{(n_{a} \ell_{a})\, j_{a}^{2j_a}}^{\text{core subshells}} \; \overbrace{(n_{m} \ell_{m})\, j_{m}^{w_m-1}\; (n_{n} \ell_{n})\, j_{n}^{w_n+1}}^{\text{valence subshells}} \; \overbrace{(n_{r} \ell_{r})\, j_{r}}^{\text{virtual subshells}}$ \\
\\
$\underbrace{- \; \frac{w_m \left( \left[j_n\right]-w_n\right)} {\sqrt{\left[j_m, j_n\right]}}
 \,
\mathcal{A}\left( 0, \, m a, \, r n \right)}_{\text{from $\text{CV}_{3}$ Feynman diagram}}$ 
$- \frac{(-1)^{j_m+j_n+j_r+j_a} w_m}{\left[j_m\right]} \left\{ \underbrace{ 2 \sum_{k} \mathcal{C}\left( k, \, m a, \, n r \right)}_{\text{from $\text{CV}_{2}^{r}\text{ and CV}_{2}^{s}$ Feynman diagrams}}
+ \underbrace{ \sum_{k} \frac{1}{\left[ k \right]} \mathcal{P} \left( kk, ma, \, nr \right)}_{\text{from $\text{CV}_{1}^{s}$ Feynman diagram}}  \right\}$ \\
\\ \hline 
\end{tabular}
\end{center}
\caption{Expressions for core-valence corrections to the energy in Eq. (\ref{eq:BogEnergy}), not depending on term.}
\label{tab:Implemen_CV1}
\end{table*}

The first type contribution of CV correlations in the second-order of perturbation theory is expressed only over $\Delta \mathcal{E}_0 \left(K J \right)$ (see Table~\ref{tab:Implemen_CV1}). 
The contributions $\Delta \mathcal{F}^{k} \left( n \ell j, \; n \ell j \right)$, $\Delta \mathcal{F}^{k} \left( n \ell j, \; n' \ell' j' \right)$,
and $\Delta \mathcal{G}^{k} \left( n \ell j, \; n' \ell' j' \right)$ are equal to zero in this case. 
The contributions $\Delta \mathcal{E}_0 \left(K J \right)$ can be expressed through $\mathcal{A}$ and $\mathcal{C}$ coefficients (see Table~\ref{tab:Implemen_CV1}) which has following expressions
\begin{equation}
\label{eq:BogA}
   \mathcal{A}\left(x, \; i j, \; i' j'\right) 
  = \sum_{k,k'}
		  \left\{
    \begin{array}{ccc}
      k  & k' & x \\
      j_{i} & j_{i} & j_{i'}
    \end{array} \right\}
				  \left\{
    \begin{array}{ccc}
      k  & k' & x \\
      j_{j'} & j_{j'} & j_{j}
    \end{array} \right\}
\mathcal{P}\left(kk', \; i j, \; i' j'\right) ,
\end{equation}

\begin{equation}
\label{eq:BogC}
   \mathcal{C}\left(k, \; i j, \; i' j'\right) 
  = \sum_{k'}
		  \left\{
    \begin{array}{ccc}
      k  & j_{i} & j_{i'} \\
      k' & j_{j} & j_{j'}
    \end{array} \right\}
\mathcal{Q}\left(kk', \; i j, \; i' j'\right) ,
\end{equation}
where
\begin{equation}
\label{eq:BogP}
   \mathcal{P}\left(kk', \; i j, \; i' j'\right) 
   = \mathcal{R}^{k}\left(i j, \; i' j'\right) \; \mathcal{R}^{k'}\left(i' j', \; i j \right)  \; 
	\mathcal{O}\left(K', K \right),
\end{equation}

\begin{equation}
\label{eq:BogQ}
   \mathcal{Q}\left(kk', \; i j, \; i' j'\right)
   = \mathcal{R}^{k}\left(i j, \; i' j'\right) \; \mathcal{R}^{k'}\left(i' j', \; j i\right)  \; 
\mathcal{O}\left(K', K \right) .
\end{equation}
We would like to emphasize once more that energy denominator is defined differently/opposite in the expressions of Feynman diagrams (see for example Fig.~\ref{CV_1}, Eq. (\ref{eq:BogP}), and Eq. (\ref{eq:BogQ}))

\begin{equation}
\label{eq:BogO1}
\mathcal{O}\left(K', K \right)
= \frac{1}{\overline{E}\left(K' \right)-\overline{E}\left(K\right)}.
\end{equation}

\begin{table*}
\begin{center}
\begin{tabular}{|lll|} \hline
Corrections & Slater integral & $k$ values\\ \hline  \hline
& & \\
$\underbrace{ 
\left[ k \right] \mathcal{Y}\left( 1, \, k, \, m a, \, r n, \, m n \right)}_{\text{from $\text{CV}_3$ Feynman diagram}}$ 
&$\Delta \mathcal{F}^{k}(m,n)$ & $k>0$\\
& & \\
$\underbrace{ 2 \left( -1 \right)^{j_r+j_a+k}  \mathcal{Z}\left(1, k, \, m a, \, n r, \; m n \right) }_{\text{from $\text{CV}_4$ and $\text{CV}_5$ Feynman diagrams}} + \underbrace{ \frac{\left( -1 \right)^{j_r+j_a+k}}{\left[ k \right] } \mathcal{P}\left( kk, \, m a, \, n r  \right)}_{\text{from $\text{CV}_6$ Feynman diagram}}$ 
&$\Delta \mathcal{G}^{k}(m,n)$ & $k \geq 0$ \\ 
& &  \\ \hline 
\end{tabular}
\end{center}
\caption{Expressions for Slater integrals $\Delta \mathcal{F}^{k}(m,n)$ and $\Delta \mathcal{G}^{k}(m,n)$ (see Eq. (\ref{eq:BogEnergy})) corresponding to the second type of core-valence 
$(n_{a} \ell_{a})\, j_{a}^{2j_a+1} \, (n_{m'} \ell_{m'})\, j_{m'}^{w_m'} \, (n_{n'} \ell_{n'})\, j_{n'}^{w_n'} 
   \rightarrow (n_{a} \ell_{a})\, j_{a}^{2j_a} \; (n_{m'} \ell_{m'})\, j_{m'}^{w_m'-1} \; (n_{n'} \ell_{n'})\, j_{n'}^{w_n'+1}
	 \; (n_{r} \ell_{r})\, j_{r}$  correlations.} 
\label{tab:Implemen_CV2}
\end{table*}

Second type contribution of CV correlations in the second-order of perturbation theory is expressed over $\Delta \mathcal{E}_0 \left(K J \right)$, $\Delta \mathcal{F}^{k} \left( n \ell j, \; n \ell j \right)$, $\Delta \mathcal{F}^{k} \left( n \ell j, \; n' \ell' j' \right)$,
and $\Delta \mathcal{G}^{k} \left( n \ell j, \; n' \ell' j' \right)$ (see 
Tables~\ref{tab:Implemen_CV1} and \ref{tab:Implemen_CV2}).

The Feynman diagram with the notation CV$_1^s$ in the Table~\ref{tab:Implemen_CV2}
means the same Feynman diagram CV$_1$ (see Fig. \ref{CV_1}) but instead of
index $s$ belonging to $G$ type of orbitals is used the index $n$ which belong to
$F'$ type of orbitals.
Similar is with the notations CV$_2^r$ and CV$_2^{s}$ in the Table~\ref{tab:Implemen_CV2}. Here index $r$ (index $s$) is changed to index $n$ (index $n$) in the diagram
CV$_2^r$ (CV$_2^{s}$). These type of Feynman diagram are coming from
reordering operators of second quantization from coupled by pairs 
$[a^{(j)} \times \tilde{a}^{(j')}]^{(k)}$
 order to normal order of operators of second quantization. 
The diagram CV$_1^s$ is coming after reorder the diagram CV$_6$, 
 the diagram CV$_2^r$ is coming after reorder the diagram CV$_4$, and the CV$_2^{s}$
is coming after reorder the diagram CV$_5$. This reordering allow us to have the same spin-angular part of the diagrams 
CV$_4$, CV$_5$, and CV$_6$ as has any two-particle operator in the {\sc Grasp}.

\begin{equation}
\label{eq:BogY}
   \mathcal{Y}\left(I, \; x, \; i j, \; i' j', \; i'' j''\right) 
  = \sum_{k,k'}
		\left( -1 \right)^{k+k'+x}
				  \left\{
    \begin{array}{ccc}
      k  & k' & x \\
      j_{i''} & j_{i} & j_{i'}
    \end{array} \right\}				  \left\{
    \begin{array}{ccc}
      k  & k' & x \\
      j_{j''} & j_{j'} & j_{j}
    \end{array} \right\}
\mathcal{T}\left(I, \; kk', \; i j, \; i' j', \; i'' j''\right) ,
\end{equation}

\begin{equation}
\label{eq:BogT}
   \mathcal{T}\left(I, \; kk', \; i j, \; i' j', \; i'' j''\right) 
  =
\left\{
   \begin{array}{ll}
\mathcal{R}^{k}\left(i j, \; i' j'\right)  \; 
\mathcal{R}^{k'}\left(i' j'', \; i'' j \right)  \; 
\mathcal{O}\left(K', K\right)
& \mbox{ for } I = 1, \\ \\
\mathcal{R}^{k}\left(i j, \; i' j'\right)  \; 
\mathcal{R}^{k'}\left(i' j'', \; i'' j \right)  \; 
\mathcal{O}\left(K', K_1 K_2 \right)
& \mbox{ for } I = 2,
   \end{array}
   \right.
\end{equation}

where
\begin{equation}
\label{eq:BogO2}
\mathcal{O}\left(K', K_1 K_2 \right)
=
\frac{1}{2} \left( {\frac{1}{\overline{E}\left(K' \right)-\overline{E}\left(K_1\right)}}
+{\frac{1}{\overline{E}\left(K' \right)-\overline{E}\left(K_2\right)}}
\right).
\end{equation}

\begin{equation}
\label{eq:BogZ}
   \mathcal{Z}\left(I, \; k, \; i j, \; i' j', \; i'' j''\right) 
  = \sum_{k'}
				  \left\{
    \begin{array}{ccc}
      k & j_{i''} & j_{j''} \\
      k' & j_{j} & j_{j'}
    \end{array} \right\}
\mathcal{U}\left(I, \; kk', \; i j, \; i' j', \; i'' j''\right) ,
\end{equation}
\begin{equation}
\label{eq:BogU}
   \mathcal{U}\left(I, \; kk', \; i j, \; i' j', \; i'' j''\right)  
	= \left\{
   \begin{array}{ll}
   \mathcal{R}^{k}\left(i j, \; i' j'\right)  \; 
\mathcal{R}^{k'}\left(j'' j', \; j i'' \right)  \; 
\mathcal{O}\left(K', K\right)
& \mbox{ for } I = 1, \\ \\
\mathcal{R}^{k}\left(i' j', \;  i j \right)  \; 
\mathcal{R}^{k'}\left(i'' j, \; j' j'' \right)  \; 
\mathcal{O}\left(K', K\right)
& \mbox{ for } I = 2, \\ \\
\mathcal{R}^{k}\left(i j, \; i' j'\right)  \; 
\mathcal{R}^{k'}\left(j'' j', \; j i'' \right)  \; 
\mathcal{O}\left(K', K_1 K_2 \right)
& \mbox{ for } I = 3, \\ \\
\mathcal{R}^{k}\left(i' j', \;  i j \right)  \; 
\mathcal{R}^{k'}\left(i'' j, \; j' j'' \right)  \; 
\mathcal{O}\left(K', K_1 K_2 \right)
& \mbox{ for } I = 4.
   \end{array}
   \right.
\end{equation}

There are following symmetry for $\mathcal{Z}$ coefficients 
\begin{equation}
\label{eq:Z_summetry1}
\mathcal{Z}\left(1, \; K, i j, \; i' j' \; i i' \right)
=
\mathcal{Z}\left(2, \; K, i j, \; i' j' \; i i' \right) ,
\end{equation}
\begin{equation}
\label{eq:Z_summetry2}
\mathcal{Z}\left(3, \; K, i j, \; i' j' \; i' i \right)
=
\mathcal{Z}\left(4, \; K,  i j, \; i' j' \; i' i \right) .
\end{equation}

\begin{table*}
\begin{center}
\begin{tabular}{|lll|} \hline
Corrections & Slater integral & $k$ values\\ \hline  \hline
& & \\ 
$2 \left\{ \left[ k \right] 
\underbrace{ \mathcal{Y}\left(2, \, k, \; m a, \; r n, \; n m \right)}_{\text{from $\text{CV}_3$ Feynman diagram}} \right.$ &$\Delta \mathcal{R}^{k}(m m,nn)$ & $k \geq 0$ \\ 
& &  \\
$+ \underbrace{ \left( -1 \right)^{j_r+j_a+k} \mathcal{Z}\left(3, \; k, \; m a, \; n r, \; n m \right)}_{\text{from $\text{CV}_4$ Feynman diagram}}
+ \underbrace{ \left( -1 \right)^{j_r+j_a+k} \mathcal{Z}\left(4, \; k, \; n  a, \; m r, \; m n\right)}_{\text{from $\text{CV}_5$ Feynman diagram}}$ & & \\
& & \\ 
$\left.
+ \underbrace{ \frac{\left( -1 \right)^{j_r+j_a+k}}{\left[ k \right]} \mathcal{S}\left(kk, \; m a, \; n r, \; m n\right)}_{\text{from $\text{CV}_6$ Feynman diagram}}
\right\}$ & & \\
& & \\ \hline
\end{tabular}
\end{center}
\caption{Expressions for Slater integral $\Delta \mathcal{R}^{k}(m m,nn)$  (see Eq. (\ref{eq:BogEnergy})) corresponding to the
core-valence \\ 
$(n_{a} \ell_{a})\, j_{a}^{2j_a+1} \, (n_{m'} \ell_{m'})\, j_{m'}^{w_m'} \, (n_{n'} \ell_{n'})\, j_{n'}^{w_n'} 
   \rightarrow (n_{a} \ell_{a})\, j_{a}^{2j_a} \; (n_{m'} \ell_{m'})\, j_{m'}^{w_m'-1} \; (n_{n'} \ell_{n'})\, j_{n'}^{w_n'+1}
	 \; (n_{r} \ell_{r})\, j_{r}$  correlations
 coming from the off diagonal matrix element 
$\redmem{(n_{m} \ell_{m})\, j_{m}^{w_m} \, (n_{n} \ell_{n})\, j_{n}^{w_n}}{\, \widehat{{\cal H}}^{(2)}_{Effective} \,}{(n_{m} \ell_{m})\, j_{m}^{w_m-2} \, (n_{n} \ell_{n})\, j_{n}^{w_n+2} }$.} 
\label{tab:Implemen_CV_Off}
\end{table*}

The contribution of CV correlation in the second-order of perturbation theory coming from off diagonal matrix element 
$\redmem{(n_{m} \ell_{m})\, j_{m}^{w_m} \, (n_{n} \ell_{n})\, j_{n}^{w_n}}{\, \widehat{{\cal H}}^{(2)}_{Effective} \,}{(n_{m} \ell_{m})\, j_{m}^{w_m-2} \, (n_{n} \ell_{n})\, j_{n}^{w_n+2} }$
are described by the diagrams CV$_3$, CV$_4$, CV$_5$, and CV$_6$. Reformulation expressions of these diagrams to the form suitable to the {\sc Grasp} gave these corrections only to the radial integral $\Delta \mathcal{R}^{k}(m m,nn)$ (see Table~\ref{tab:Implemen_CV_Off}). All these formulas are expressed via the quantities introduced earlier in the paper except the: 

\begin{equation}
\label{eq:BogS}
  \mathcal{S}\left(kk', \; i j, \; i' j', \; i'' j''\right)  
  = 
\mathcal{R}^{k}\left(i j, \; i' j' \right)  \; 
\mathcal{R}^{k'}\left(j' i'', \; j j'' \right) \; 
\mathcal{O}\left(K', K_1 K_2 \right).
\end{equation}

This theory in irreducible tensorial form is more suitable to be included in such version of the {\sc Grasp} which is based on configuration state function generators
(CSFGs)~\cite{grasp2023}. This is related to the fact that this version of the package allows us to distinguish $F$, $F'$, and $G$ sets of orbitals very easily in the process of computing atomic data. In the following section we will present a test case of this implementation.

\section{Calculation core-valence correlations with new implementation}
This section is intended to present the results using here newly developed and presented method,
based on the Rayleigh-Schr\"odinger perturbation theory in irreducible tensorial form, which is implemented in the {\sc Grasp}2018 package.
This new developed method allows us to include the CV correlations in the computations choosing the preferred core and virtual orbitals.
Using this method we can open the core fully and estimate the impact of CV correlations
in the second order of perturbation theory for the computed levels. 
To calculate the contributions of such CV correlations the average energies of the configurations are needed.
There are three options to calculate average energy of the configuration in the program implemented in the {\sc Grasp}2018 package:
\begin{itemize}
	\item The average energy is calculated using an analytical expression (Eq. (\ref{eq:simple_average})) (option 0);
  \item The average energy is calculated according to diagonal matrix element (option 1);
  \item The average energy is calculated by discarding those energies for which the non-diagonal matrix elements with multi-reference set are zero (option 2).
\end{itemize}
Results from the computations using the new method will be marked as \textbf{CV RCI+RSMBPT} (accordig to Eq. (\ref{eq:BogEnergy})).
To evaluate the accuracy and the reliability of the results, these are compared with the results from rearranged RCI method, 
based on CSFGs~\cite{grasp2023}, which compute only the CV correlations (results marked as \textbf{CV RCI}). 

\subsection{First test case}
\label{sec:PT_1_Test_Case}

In first test 3 energy levels of the ground ($\mathrm{1s^22s^22p^63s^23p^3}$) configuration with $J$=3/2 of the Cl III ion are computed. The MR set consists of 
3 CSFs ($\mathrm{3s^23p^3}$, $\mathrm{3s^23p_{-}3p^2}$, $\mathrm{3s^23p^2_{-}3p}$ in $jj$-coupling) belonging to $\mathrm{3s^23p^3}$configuration from which substitutions are allowed to include CV correlations.
The 1s, 2s, $\mathrm{2p_-}$ and 2p subshells are defined as opened core subshells ($F$ set), 3s, $\mathrm{3p_-}$, and 3p as valence subshells ($F'$ set), and 4s, $\mathrm{4p_-}$, 4p, $\mathrm{3d_-}$, 3d as virtual ones ($G$ set). 
Such set of virtual orbitals will be marked as L1.
So the CSFs list consists of 3 CSFs belonging to MR set, the rest CSFs are CV correlations of the MR set. Radial wavefunctions are taken from the earlier computations \cite{Rynkun_2019}.

Firstly, we will check the expressions derived in sections \ref{sec:seconOrder} and \ref{sec:PT_implementation}.
As it is described in section \ref{sec:PT_implementation} 
there are implemented few types of CV correlations.
Below the contributions for each type of the CV correlations will be 
compared with the results from CV RCI computations.
To calculate the contribution of particular $K'$ configuration of the CV correlations from the RCI, based on the CSFGs, 
for the first and second type (see Tables \ref{tab:Implemen_CV1} and \ref{tab:Implemen_CV2}) we will use such equation: 
\begin{eqnarray}
\label{eq:H_E}
\Delta E_{PT} = - \frac{\sum_{\chi'} \left| \redmem{\left(K \chi J \right)}{\, V \,}{\left(K' \chi' J \right)}\right|^2}{\overline{E}\left(K' \right)-\overline{E}\left(K\right)}
\end{eqnarray}
and 
\begin{eqnarray}
\label{eq:off_diagonal}
\Delta E_{PT} = - \frac{\sum_{\chi'}  \redmem{\left(K_1 \chi J \right)}{\, V \,}{\left(K' \chi' J \right)} \redmem{\left(K_2 \chi J \right)}{\, V \,}{\left(K' \chi' J \right)}}{2} 
\nonumber \\
	   \times \left( {\frac{1}{\overline{E}\left(K' \right)-\overline{E}\left(K_1\right)}}
+{\frac{1}{\overline{E}\left(K' \right)-\overline{E}\left(K_2\right)}}
\right)
\end{eqnarray}
to compute the impact from off-diagonal matrix elements (see Table \ref{tab:Implemen_CV_Off}).

The results of these examples for the first and second type of CV correlations are presented in Table \ref{check_H_E}.
Matrix elements and average energies (with option 0) are given in the Table. The CV contribution from CV RCI+RSMBPT computations is compared with contribution calculated by Eq. (\ref{eq:H_E}).
As seen from the Table, there is an excellent agreement between two computations.
For the first type the impact of CV correlations is
-2.48300702640003E-04 a.u. by Eq. (\ref{eq:H_E}), and
-2.48300702643986E-04 a.u. by Eq. (\ref{eq:BogEnergy_PT}); for the second type of the CV correlations 
-5.30286558582154E-05 a.u. by Eq. (\ref{eq:H_E}), and 
-5.30286558628605E-05 a.u. by Eq. (\ref{eq:BogEnergy_PT}).
The example for the contribution of CV correlations to off-diagonal matrix elements is presented in Table \ref{check_off_diadonal}.
It is seen, that there is an excellent agreement between both calculations (7.74908996437684E-06 a.u. by Eq. (\ref{eq:off_diagonal}) and 7.74908996512910E-06 a.u. by Eq. (\ref{eq:BogEnergy_PT})), too.
By using the Rayleigh-Schr\"odinger perturbation theory in irreducible tensorial form to calculate the contribution of CV correlations
14 numbers after comma are reproduced.

\begin{table*}[!ht]
\setlength{\tabcolsep}{2pt}
{\footnotesize
\caption{The comparison of the contribution of the first and second type of CV correlations from the CV RCI and CV RCI+RSMBPT computations. 
The contribution of the $K'$ CV correlations is given for the computed level $\mathrm{1s^22s^22p^2_{-}2p^43s^23p^3}$.
}            
\label{check_H_E}
\centering
\begin{tabular}{l r r}
\hline\hline
\multicolumn{1}{c}{$K' \chi' J=3/2$}&\multicolumn{1}{c}{$\redmem{\left(K \chi J=3/2 \right)}{\, V \,}{\left(K' \chi' J=3/2 \right)}$} & \multicolumn{1}{c}{Energy, a.u.}  \\
\hline
\noalign{\smallskip}
\multicolumn{3}{c}{$\mathrm{1s^22s^22p_{-}2p^43s^23p^23d_{-}3d}$ (first type of CV (see section \ref{subsec:first_type}))} \\
 $\mathrm{1s^22s^22p_{-}2p^43s^23p^2(0)3d_{-}<1>3d}$& 1.9323531771818409E-02 & $\overline{E}\left(K\right)$=-459.55556158934053  \\
 $\mathrm{1s^22s^22p_{-}2p^43s^23p^2(0)3d_{-}<2>3d}$& 1.1447066004165937E-03 & $\overline{E}\left(K'\right)$=-450.50099438062546 \\
 $\mathrm{1s^22s^22p_{-}2p^43s^23p^2(2)<3/2>3d_{-}<1>3d}$&-1.0237704805308916E-03 \\
 $\mathrm{1s^22s^22p_{-}2p^43s^23p^2(2)<3/2>3d_{-}<2>3d}$& 1.0152134678921704E-02 \\
 $\mathrm{1s^22s^22p_{-}2p^43s^23p^2(2)<3/2>3d_{-}<3>3d}$&-2.8273270317335265E-02 \\
 $\mathrm{1s^22s^22p_{-}2p^43s^23p^2(2)<5/2>3d_{-}<1>3d}$&-6.0999204304291775E-03 \\
 $\mathrm{1s^22s^22p_{-}2p^43s^23p^2(2)<5/2>3d_{-}<2>3d}$& 1.8300440014309707E-02 \\
 $\mathrm{1s^22s^22p_{-}2p^43s^23p^2(2)<5/2>3d_{-}<3>3d}$&-2.4406821873536943E-02 \\
 $\mathrm{1s^22s^22p_{-}2p^43s^23p^2(2)<5/2>3d_{-}<4>3d}$&-1.4987727556927458E-03 \\
\\
\multicolumn{3}{l}{$\Delta E_{PT}$=-2.48300702640003E-04 a.u. (by Eq. (\ref{eq:H_E}))}\\
\multicolumn{3}{l}{$\Delta E_{PT}$=-2.48300702643986E-04 a.u. (by Eq. (\ref{eq:BogEnergy_PT}))}\\
\hline
\noalign{\smallskip}
\multicolumn{3}{c}{$\mathrm{1s^22s^22p_{-}2p^43s3p_{-}3p^34s}$ (second type of CV (see section \ref{subsec:second_type}))} \\ 
$\mathrm{1s^22s^22p_{-}2p^43s<0>3p_{-}<1/2>3p^3<1>4s}$&7.9640894479663035E-03 & $\overline{E}\left(K\right)$=-459.55556158934053 \\
$\mathrm{1s^22s^22p_{-}2p^43s<0>3p_{-}<1/2>3p^3<2>4s}$&1.0281595266559909E-02 & $\overline{E}\left(K'\right)$=-450.58662000712661\\
$\mathrm{1s^22s^22p_{-}2p^43s<1>3p_{-}<1/2>3p^3<1>4s}$&1.0720418507720091E-02 \\
$\mathrm{1s^22s^22p_{-}2p^43s<1>3p_{-}<1/2>3p^3<2>4s}$&1.3840000781591226E-02 \\
$\mathrm{1s^22s^22p_{-}2p^43s<1>3p_{-}<3/2>3p^3<1>4s}$&0.0 \\
$\mathrm{1s^22s^22p_{-}2p^43s<1>3p_{-}<3/2>3p^3<2>4s}$&0.0 \\
\\
\multicolumn{3}{l}{$\Delta E_{PT}$=-5.30286558582154E-05 a.u. (by Eq. (\ref{eq:H_E}))}\\
\multicolumn{3}{l}{$\Delta E_{PT}$=-5.30286558628605E-05 a.u. (by Eq. (\ref{eq:BogEnergy_PT}))}\\
\hline
\hline
\end{tabular}
}
\end{table*}

\begin{table*}[!ht]
\setlength{\tabcolsep}{2pt}
{\footnotesize
\caption{The comparison of the contribution of CV correlations to off-diagonal matrix elements from the CV RCI and CV RCI+RSMBPT computations.
The contribution of the $K'$ CV correlations is given to the off-diagonal matrix elements between computed levels
$\mathrm{1s^22s^22p^2_{-}2p^43s^23p^3}$ ($K_1$) and $\mathrm{1s^22s^22p^2_{-}2p^43s^23p^2_{-}3p}$ ($K_2$).
}            
\label{check_off_diadonal}
\centering
\begin{tabular}{l r r}
\hline\hline
\multicolumn{1}{c}{$K' \chi' J=3/2$}&
\multicolumn{1}{c}{$\redmem{\left(K_1 \chi J=3/2 \right)}{\, V \,}{\left(K' \chi' J=3/2 \right)}$} & 
\multicolumn{1}{c}{$\redmem{\left(K_2 \chi J=3/2 \right)}{\, V \,}{\left(K' \chi' J=3/2 \right)}$} \\
\hline
\noalign{\smallskip}
\multicolumn{3}{c}{$\mathrm{1s^22s2p_{-}^22p^43s^23p_{-}3p^23d}$ (off-diagonal (see section \ref{subsec:offdiagonal_type})} \\
$\mathrm{1s^22s2p_{-}^22p^43s^23p_{-}<1>3p^2(0)3d}$& 2.5856710447641450E-03 & 8.0873774303662094E-03  \\
$\mathrm{1s^22s2p_{-}^22p^43s^23p_{-}<0>3p^2(2)<2>3d}$& 0.0 & 0.0  \\
$\mathrm{1s^22s2p_{-}^22p^43s^23p_{-}<1>3p^2(2)<1>3d}$& -8.1766097814018907E-04 & 2.5574532977397013E-03 \\
$\mathrm{1s^22s2p_{-}^22p^43s^23p_{-}<1>3p^2(2)<2>3d}$& -2.7928439287496273E-03 & 8.7353660093932462E-03 \\
$\mathrm{1s^22s2p_{-}^22p^43s^23p_{-}<1>3p^2(2)<3>3d}$& -4.9959911001853999E-03 & 1.5626297764275494E-02 \\
\\
\multicolumn{3}{l}{$\overline{E}\left(K_1\right)$=-459.55556158934053 a.u.} \\
\multicolumn{3}{l}{$\overline{E}\left(K_2\right)$=-459.56656073844789 a.u.} \\
\multicolumn{3}{l}{$\overline{E}\left(K'\right)$=-448.76686304559024 a.u.} \\
\\
\multicolumn{3}{l}{$\Delta E_{PT}$=7.74908996437684E-06 a.u. (by Eq. (\ref{eq:off_diagonal}))} \\
\multicolumn{3}{l}{$\Delta E_{PT}$=7.74908996512910E-06 a.u. (by Eq. (\ref{eq:BogEnergy_PT}))} \\
\hline
\hline
\end{tabular}
}
\end{table*}

\begin{table*}[!ht]
\setlength{\tabcolsep}{2pt}
{\scriptsize
\caption{The energy levels (in cm$^{-1}$) and total energies (in a.u.) for 3 energy levels of $\mathrm{3s^23p^3}$ configuration 
with $J$=3/2 are given from both computations when CV correlations are included. 
0, 1, and 2 columns mark 
the options of averaging the energy of the configuration. 
}            
\label{comp}
\centering
\begin{tabular}{r l r r r r r r r}
\hline\hline
\multicolumn{1}{c}{No.} & \multicolumn{1}{c}{State} & \multicolumn{1}{c}{MR}& \multicolumn{2}{c}{CV RCI}&& \multicolumn{3}{c}{CV RCI+RSMBPT} \\
\cline{4-5} \cline{7-9}
&&&  & \multicolumn{1}{c}{ZF} && \multicolumn{1}{c}{0} & \multicolumn{1}{c}{1} & \multicolumn{1}{c}{2} \\
\hline
\noalign{\smallskip}
&&\multicolumn{7}{c}{Energy evels (cm$^{-1}$)} \\
1& $\mathrm{3s^23p^3(^4_3S)~^4S_{3/2}}$&     0.00&     0.00&     0.00&&     0.00&     0.00&     0.00 \\                        
2& $\mathrm{3s^23p^3(^2_3D)~^2D_{3/2}}$& 23142.25& 23141.15& 23136.12&& 23140.51& 23140.23& 23140.32 \\                        
3& $\mathrm{3s^23p^3(^2_1P)~^2P_{3/2}}$& 38735.68& 38732.59& 38723.48&& 38730.87& 38729.48& 38729.83 \\                        
\\                                                                                                                             
&&\multicolumn{7}{c}{Total energy (a.u.)} \\                                                                                    
1& $\mathrm{3s^23p^3(^4_3S)~^4S_{3/2}}$& -459.6785245& -459.6807614& -459.6807760&& -459.6808103& -459.6807921& -459.6807918 \\
2& $\mathrm{3s^23p^3(^2_3D)~^2D_{3/2}}$& -459.5730806& -459.5753225& -459.5753600&& -459.5753744& -459.5753574& -459.5753568 \\
3& $\mathrm{3s^23p^3(^2_1P)~^2P_{3/2}}$& -459.5020317& -459.5042827& -459.5043388&& -459.5043395& -459.5043276& -459.5043257 \\
\hline
\noalign{\smallskip}
\multicolumn{2}{r}{$N_{CSFs}$} & 3& 1941& 1941&& 3& 3& 3 \\
\hline
\hline
\end{tabular}
}
\end{table*}

The energy structure results from both (CV RCI and CV RCI+RSMBPT) computations are given in Table \ref{comp}. 
For CV RCI case in the RCI computations also ZF method is applied. The MR space is used as the principal part $P$.
In CV RCI+RSMBPT case results from three computations (different options of averaging energies) are presented in the Table.
As is seen from Table \ref{comp} the differences between CV RCI and CV RCI+RSMBPT (option 0)
calculations are 0.0000489 a.u., 0.0000519 a.u., and 0.0000568 a.u., 
whereas when the contributions of CV correlations are -0.0022369 a.u., -0.0022419 a.u., and -0.0022510 a.u., 
respectively for the $\mathrm{^4S_{3/2}}$, $\mathrm{^2D_{3/2}}$, and $\mathrm{^2P_{3/2}}$ levels.
The disagreement between the results in this case is only 2.2-2.5\%.
This disagreement decreases when in CV RCI+RSMBPT computations energy of configuration 
is averaged according to diagonal matrix element (option 1) 1.4-2.0\%.
There is a little decrease when average energy is calculated by discarding those energies 
for which the non-diagonal matrix elements with multi-reference set are zero (option 2).
Comparing CV RCI+RSMBPT with results from CV RCI ZF this disagreement decreases till 0.1-0.7\% in option 2.
We want to note, that the difference between CV RCI and CV RCI+RSMBPT computations 
for 3 computed levels is similar and stable (about 0.00005 a.u with option (0)).
Meanwhile comparing CV RCI+RSMBPT results with results from CV RCI ZF 
these differences differs for each computed level, and are 0.0000343 a.u., 0.0000144 a.u., and 0.0000007 a.u., respectively for first, second, and third level.
Comparing the splitting of the computed levels from CV RCI and CV RCI+RSMBPT computations, we see, 
that splitting from CV RCI+RSMBPT is closer to the CV RCI results than CV RCI ZF.

The obtained results and comparisons made lead us to the following, that Rayleigh-Schr\"odinger perturbation theory in irreducible tensorial form expressions derived in sections \ref{sec:seconOrder} and \ref{sec:PT_implementation} are correct, the assumptions made in section \ref{sec:PT_implementation} are not materially wrong,
and the program based on this methodology is free from bugs. Thus it mean, that the direct inclusion of correlations with the Brillouin-Wigner perturbation theory 
from subsection~\ref{sec:ZFOm} can be substituted for by 
the version of the Rayleigh-Schr\"odinger perturbation theory presented in the subsection~\ref{sec:seconOrder}.
It is appropriate because by using the combination of the RCI and the Rayleigh-Schr\"odinger perturbation theory in irreducible tensorial form, the CSFs space consist only from 3 CSFs in CV RCI+RSMBPT computations comparing to 1941 CSFs that are in the ordinary RCI calculations (CV RCI) and in the CV RCI ZF.
Such significant CSFs reductions reduce the matrix and CPU computing time, too.

\subsection{Second test case}
\label{sec:PT_2_Test_Case}


In this case the same test as described above is chosen.
We use the program which is designed to determine the contribution 
of each $K'$ configuration of CV correlations for CSF for which energy need to be calculated according to 
Rayleigh-Schr\"odinger perturbation theory in irreducible tensorial form according to the Eq. (\ref{eq:BogEnergy_PT}).
The program gives the total contribution of CV correlations with selected core and virtual orbitals, 
and calculates the contribution of each $K'$ configuration of the CV correlations for the computed levels.
$K'$ configurations are sorted in descending order according to the impact of the CV correlations for each level.
Further, we select $K'$ configurations by the CV correlations impact
with the specified fraction of the total CV contribution, and perform RCI computations including them. These fractions are presented in the percentage.
Results from the computations when the CV correlations are included using the method described above are marked as \textbf{CV RCI (RSMBPT)}, 
and usual RCI calculations (when all CV correlations are included) are marked as \textbf{CV RCI} and are given in Table \ref{comp3}. There are also given energies without 
CV correlations (marked as MR).
In the both (CV RCI and CV RCI (RSMBPT)) computations also ZF method is applied (the MR space is used as the principal part $P$).

As seen from the Table, by including step by step the most important $K'$ configurations of the CV correlations, 
the results smoothly converge to the CV RCI computations. 
In the case when 99.995\% of CV correlations are included in the computations we reproduce the results of usual RCI computations.
The convergence by including the most important $K'$ configurations of the CV correlations for computed 3 energy levels 
is displayed in Figure \ref{convergence}. The trend of inclusion the most important CV correlations in ZF case is very similar.
From Table \ref{comp3} it seen, that in case when 99.995\% of CV correlations are included the CSFs space decreases about 27\% comparing to the space in CV RCI computations. By using Rayleigh-Schr\"odinger perturbation theory in irreducible tensorial form in such way, we can collect the most important CV correlations, thus the CSFs space is decreased (the matrix and the computing CPU time is reduced).

\begin{table*}[!ht]
\setlength{\tabcolsep}{2pt}
{\scriptsize
\caption{The energy levels (in cm$^{-1}$) and total energies (in a.u.) for 3 energy levels of $\mathrm{3s^23p^3}$ configuration 
with $J$=3/2 are given from both computations when CV correlations are included. (L1)}            
\label{comp3}
\centering
\begin{tabular}{r l r r r r r r r r}
\hline\hline
\multicolumn{1}{c}{No.} & \multicolumn{1}{c}{State} & \multicolumn{1}{c}{MR}& \multicolumn{1}{c}{CV RCI}& \multicolumn{6}{c}{CV RCI (RSMBPT)} \\
\cline{5-10}
&&&& \multicolumn{1}{c}{95\%} & \multicolumn{1}{c}{99\%} & \multicolumn{1}{c}{99.5\%} & \multicolumn{1}{c}{99.95\%} & \multicolumn{1}{c}{99.995\%} & \multicolumn{1}{c}{99.999999\%} \\
\hline
\noalign{\smallskip}
&&\multicolumn{8}{c}{Energy levels (cm$^{-1}$)} \\
1& $\mathrm{3s^23p^3(^4_3S)~^4S_{3/2}}$&     0.00&     0.00&     0.00&     0.00&     0.00&     0.00&     0.00&     0.00 \\                            
2& $\mathrm{3s^23p^3(^2_3D)~^2D_{3/2}}$& 23142.25& 23141.15& 23141.71& 23141.27& 23141.18& 23141.15& 23141.15& 23141.15 \\                            
3& $\mathrm{3s^23p^3(^2_1P)~^2P_{3/2}}$& 38735.68& 38732.59& 38734.73& 38733.04& 38732.78& 38732.62& 38732.61& 38732.61 \\                            
\\                                                                                                                                          
&&\multicolumn{8}{c}{Total energy (a.u.)} \\                                                                                                 
1& $\mathrm{3s^23p^3(^4_3S)~^4S_{3/2}}$& -459.6785245& -459.6807614& -459.6806652& -459.6807441& -459.6807518& -459.6807602& -459.6807612& -459.6807613 \\
2& $\mathrm{3s^23p^3(^2_3D)~^2D_{3/2}}$& -459.5730806& -459.5753225& -459.5752238& -459.5753047& -459.5753129& -459.5753214& -459.5753223& -459.5753224 \\
3& $\mathrm{3s^23p^3(^2_1P)~^2P_{3/2}}$& -459.5020317& -459.5042827& -459.5041768& -459.5042634& -459.5042723& -459.5042814& -459.5042824& -459.5042825 \\
\hline
\noalign{\smallskip}
\multicolumn{10}{c}{ZF} \\
&&\multicolumn{8}{c}{Energy levels (cm$^{-1}$)} \\
1& $\mathrm{3s^23p^3(^4_3S)~^4S_{3/2}}$&     0.00&     0.00&     0.00&     0.00&     0.00&     0.00&     0.00&     0.00 \\
2& $\mathrm{3s^23p^3(^2_3D)~^2D_{3/2}}$& 23142.25& 23136.12& 23136.75& 23136.28& 23136.18& 23136.13& 23136.12& 23136.12 \\
3& $\mathrm{3s^23p^3(^2_1P)~^2P_{3/2}}$& 38735.68& 38723.48& 38725.86& 38723.97& 38723.69& 38723.47& 38723.48& 38723.48 \\
\\
&&\multicolumn{8}{c}{Total energy (a.u.)} \\
1& $\mathrm{3s^23p^3(^4_3S)~^4S_{3/2}}$& -459.6785245& -459.6807760& -459.6806803& -459.6807590& -459.6807667& -459.6807749& -459.6807758& -459.6807760 \\
2& $\mathrm{3s^23p^3(^2_3D)~^2D_{3/2}}$& -459.5730806& -459.5753600& -459.5752615& -459.5753423& -459.5753505& -459.5753590& -459.5753599& -459.5753600 \\
3& $\mathrm{3s^23p^3(^2_1P)~^2P_{3/2}}$& -459.5020317& -459.5043388& -459.5042323& -459.5043196& -459.5043286& -459.5043378& -459.5043387& -459.5043388 \\
\hline
\noalign{\smallskip}
\multicolumn{2}{r}{$N_{CSFs}$} & 3& 1941& 678& 977& 1088& 1263& 1400& 1609 \\
\hline
\hline
\end{tabular}
}
\end{table*}

\begin{figure}
\centering
\includegraphics[width=\hsize]{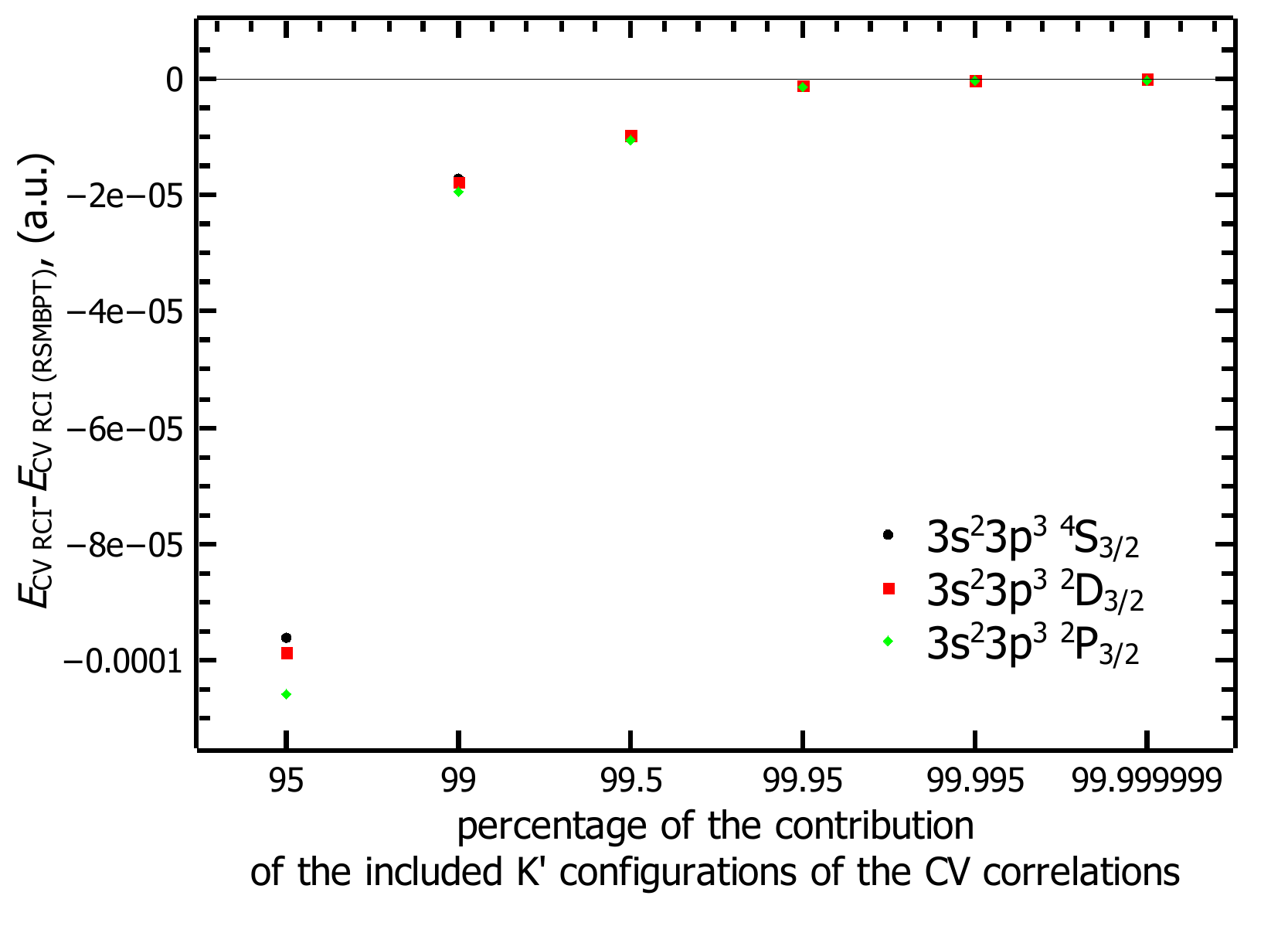}
\caption{\label{convergence} The convergence by including the most important $K'$ configurations of the CV correlations for computed 3 energy levels of $\mathrm{3s^23p^3}$ configuration with $J$=3/2. }
\end{figure}

We will take one more example to present the results of CV correlations from CV RCI and CV RCI (RSMBPT). 
We will change the 
virtual orbitals to the highest ones (9s, $\mathrm{9p_-}$, 9p, $\mathrm{9d_-}$, 9d, $\mathrm{8f_-}$, 8f, $\mathrm{8g_-}$, 8g, $\mathrm{8h_-}$, 8h, $\mathrm{8i_-}$, 8i (set of these virtual orbitals will be marked as L5)) in the test case described above.
The results are presented in Table \ref{comp3_L5}. It is seen, that the results from CV RCI (RSMBPT) smoothly converge to the CV RCI computations.
In case when 99.95\% of CV correlations are included in the computations, the results are very close to the CV RCI computations, 
and in case 99.995\% the results of usual RCI computations are reproduced.
When 99.995\% of CV correlations are included, the CSFs space 
decreases about 47\% comparing to the space in CV RCI computations.
Comparing the results from this test with previous test (with smaller virtual orbitals)
it is seen, that the contribution of CV correlations decreases (as it is known).
However using Rayleigh-Schr\"odinger perturbation theory in irreducible tensorial form in such way, 
we can calculate and estimate the contribution of each $K'$ of the CV correlations with preferred core and virtual orbitals.
That allows us to select the most important CV correlations and perform the RCI computations in the space of CSF which is significantly reduced.

\begin{table*}[!ht]
\setlength{\tabcolsep}{2pt}
{\scriptsize
\caption{The energy levels (in cm$^{-1}$) and total energies (in a.u.) for 3 energy levels of $\mathrm{3s^23p^3}$ configuration 
with $J$=3/2 are given from both computations when CV correlations are included. (L5)}            
\label{comp3_L5}
\centering
\begin{tabular}{r l r r r r r r r r}
\hline\hline
\multicolumn{1}{c}{No.} & \multicolumn{1}{c}{State} & \multicolumn{1}{c}{MR}& \multicolumn{1}{c}{CV RCI}& \multicolumn{6}{c}{CV RCI (RSMBPT)} \\
\cline{5-10}
&&&& \multicolumn{1}{c}{95\%} & \multicolumn{1}{c}{99\%} & \multicolumn{1}{c}{99.5\%} & \multicolumn{1}{c}{99.95\%} & \multicolumn{1}{c}{99.995\%} & \multicolumn{1}{c}{99.99999\%} \\
\hline
\noalign{\smallskip}
&&\multicolumn{8}{c}{Energy levels (cm$^{-1}$)} \\
1& $\mathrm{3s^23p^3(^4_3S)~^4S_{3/2}}$&     0.00&     0.00&     0.00&     0.00&     0.00&     0.00&     0.00&     0.00 \\
2& $\mathrm{3s^23p^3(^2_3D)~^2D_{3/2}}$& 23142.25& 23141.47& 23141.84& 23141.61& 23141.55& 23141.48& 23141.48& 23141.47 \\
3& $\mathrm{3s^23p^3(^2_1P)~^2P_{3/2}}$& 38735.68& 38710.07& 38710.92& 38710.05& 38710.18& 38710.10& 38710.07& 38710.07 \\
\\
&&\multicolumn{8}{c}{Total energy (a.u.)} \\
1& $\mathrm{3s^23p^3(^4_3S)~^4S_{3/2}}$& -459.6785245& -459.6797756& -459.6797214& -459.6797669& -459.6797713& -459.6797752& -459.6797755& -459.6797756 \\
2& $\mathrm{3s^23p^3(^2_3D)~^2D_{3/2}}$& -459.5730806& -459.5743353& -459.5742794& -459.5743259& -459.5743306& -459.5743348& -459.5743352& -459.5743353 \\
3& $\mathrm{3s^23p^3(^2_1P)~^2P_{3/2}}$& -459.5020317& -459.5033995& -459.5033415& -459.5033909& -459.5033947& -459.5033989& -459.5033995& -459.5033995 \\
\hline
\noalign{\smallskip}
\multicolumn{10}{c}{ZF} \\
&&\multicolumn{8}{c}{Energy levels (cm$^{-1}$)} \\
1& $\mathrm{3s^23p^3(^4_3S)~^4S_{3/2}}$&     0.00&     0.00&     0.00&     0.00&     0.00&     0.00&     0.00&     0.00 \\
2& $\mathrm{3s^23p^3(^2_3D)~^2D_{3/2}}$& 23142.25& 23140.70& 23141.08& 23140.82& 23140.76& 23140.71& 23140.70& 23140.70 \\
3& $\mathrm{3s^23p^3(^2_1P)~^2P_{3/2}}$& 38735.68& 38708.85& 38709.76& 38708.84& 38708.96& 38708.89& 38708.86& 38708.85 \\
\\
&&\multicolumn{8}{c}{Total energy (a.u.)} \\
1& $\mathrm{3s^23p^3(^4_3S)~^4S_{3/2}}$& -459.6785245& -459.6797795& -459.6797259& -459.6797709& -459.6797753& -459.6797791& -459.6797795& -459.6797795 \\
2& $\mathrm{3s^23p^3(^2_3D)~^2D_{3/2}}$& -459.5730806& -459.5743428& -459.5742873& -459.5743336& -459.5743382& -459.5743423& -459.5743427& -459.5743428 \\
3& $\mathrm{3s^23p^3(^2_1P)~^2P_{3/2}}$& -459.5020317& -459.5034090& -459.5033512& -459.5034005& -459.5034043& -459.5034084& -459.5034090& -459.5034090 \\
\hline
\noalign{\smallskip}
\multicolumn{2}{r}{$N_{CSFs}$} & 3& 10396& 2897& 4317& 4648& 5136& 5509& 6182 \\
\hline
\hline
\end{tabular}
}
\end{table*}

\section{Summary and conclusions}
The Rayleigh-Schr\"odinger perturbation theory in nonrelativistic approach~\cite{Meratal:86,Gaigalas:89} was extended to relativistic approach in such way that the Feynman diagrams corresponding to core-valence correlation are presented in irreducible tensorial form in $jj$-coupling. This allows us to use the spin-angular 
library~\cite{Gaigalas_1997,Gaigalas:2022} without any modifications and to make the easiest implementation of this theory in the {\sc Grasp}2018 package~\cite{grasp2018} in
general.
Two proposed computational methods (CV RCI+RSMBPT and CV RCI (RSMBPT)), which combine the RCI method  
and the stationary second-order Rayleigh-Schr\"odinger many-body perturbation theory in irreducible tensorial form, allows us to include the core-valence correlations for atoms or ions with any number of valence electrons in a simplified way instead of the large RCI computations.
Rayleigh-Schr\"odinger perturbation theory in irreducible tensorial form has few advantages over the Brillouin-Wigner perturbation theory implemented in the {\sc Grasp} code~\cite{grasp2013,grasp2018}: 
firstly, the space of CSF is significantly reduced, which leads to smaller matrix and its simpler
diagonalization.
So the computational resources needed for computations and the computing CPU time are reduced, too.
In addition, this computational method allows us to include the CV correlations from the deeper core.



\begin{thebibliography}{10}

\bibitem{fischerBook:77}
C. Froese Fischer, {\it The Hartree-Fock Method for Atoms}, (Wiley, New York, 1997).

\bibitem{grantBook:07} 
I. P. Grant, {\it Relativistic Quantum Theory of Atoms and Molecules}, (Springer, New York, 2007).

\bibitem{LindgrenBook:82}
I. Lindgren and J. Morrison, {\it Atomic Many-Body Theory, Springer-Verlag Berlin Heidelberg}, (New York, 1982).

\bibitem{Gaigalas_1996}
G. Gaigalas and Z. Rudzikas, On the secondly quantized theory of the many-electron atom, J. Phys. B: At. Mol. Opt. Phys. {\bf 29} (15), 3303 (1996), http://doi.org/10.1088/0953-4075/29/15/007.
	
\bibitem{Gaigalas_1997}
G. Gaigalas, Z. Rudzikas, and C. Froese Fischer, An efficient approach for spin-angular integrations in atomic structure calculations, J. Phys. B: At. Mol. Opt. Phys. {\bf 30} (17), 3747 (1997), http://doi.org/10.1088/0953-4075/30/17/006.

\bibitem{Bogetal:1997}
P. Bogdanovich, G. Gaigalas,A. Momkauskait{\.e}, and Z. Rudzikas, Accounting for admixed configurations in the second order of perturbation theory for complex atoms, Phys. Scr. {\bf 56}, 230-239 (1997), https://doi.org/10.1088/0031-8949/56/3/002.

\bibitem{DzuFla:07}
V. A. Dzuba and V. V. Flambaum, Core-valence correlations for atoms with open shells, Phys. Rev. A {\bf 75}, 052504 (2007), https://doi.org/10.1103/PhysRevA.75.052504.

\bibitem{grasp2018}
C. Froese Fischer, G. Gaigalas, P. J\"onsson, and J. Biero\'n, GRASP2018 - A fortran 95 version of the General Relativistic Atomic Structure package, Comput. Phys. Commun. {\bf 237}, 184-187 (2019), https://doi.org/10.1016/j.cpc.2018.10.032.

\bibitem{Meratal:86}
G. Merkelis, G. Gaigalas, and J. G. Kaniauskas, and Z. Rudzikas, Application of the graphical
method of the Angular Momentum Theory to the Study of the Stationary Perturbation series, Izvest. Acad. Nauk SSSR, Phys. Coll. {\bf 50}, 1403–1410 (1986). [in Russian]

\bibitem{Gaigalas:89}
G. Gaigalas, {\it Irreducible Tensorial Form of the Stationary Perturbation Theory for Atoms and Ions with Open Shells}, (PhD thesis, Institute of Physics,  Vilnius, 1989). [in Russian]

\bibitem{Gaigalas:2022}
G. Gaigalas, A program library for computing pure spin-angular coefficients for one- and two-particle operators in relativistic atomic theory, Atoms {\bf 10} (4), 129 (2022), https://doi.org/10.3390/atoms10040129.

\bibitem{grasp2013}
P. J\"onsson, G. Gaigalas, J. Biero\'n, C. Froese Fischer, and I. P. Grant, New version: GRASP2K relativistic atomic structure package, Comput. Phys. Commun. {\bf 184} (9), 2197-2203 (2013), https://doi.org/10.1016/j.cpc.2013.02.016.

\bibitem{Fisetal:16a}
C. Froese Fischer, M. Godefroid, T. Brage, P. J\"onsson, and G. Gaigalas, Advanced multiconfiguration methods for complex atoms: I. Energies and wave functions, J. Phys. B: At. Mol. Opt. Phys. {\bf 49} (18), 182004 (2016), https://doi.org/10.1088/0953-4075/49/18/182004.

\bibitem{Jonatal:2023}
P. J\"onsson, M. Godefroid, G. Gaigalas, J. Ekman, J. Grumer, W. Li, J. Li, T. Brage, I. P. Grant, J. Biero\'n, and C. Froese Fischer, An introduction to relativistic theory as implemented in GRASP, Atoms {\bf 11} (1), 7 (2023), https://doi.org/10.3390/atoms11010007.

\bibitem{grasp1989}
K.G. Dyall, I.P. Grant, C.T. Johnson, F.A. Parpia, and E.P. Plummer, GRASP: A general-purpose relativistic atomic structure program, Comput. Phys. Commun. {\bf 55} (3). 425-456 (1989), https://doi.org/10.1016/0010-4655(89)90136-7.

\bibitem{Kato_2001}
D. Kato, X. M. Tong, H. Watanabe, T. Fukami, T. Kinugawa, C. Yamada, S. Ohtani, and T. Watanabe, Fine-structure in 3d$^4$ states of highly charged Ti-like ions, J. Chin. Chem. Soc. {\bf 48} (3), 525-529 (2001).
	
\bibitem{Yutetal:62a}
A.P. Yutsis, I.NB. Levinson, and V.V. Vanagas, {\it Mathematical Apparatus of the Theory of Angular Momentum}, (Israel Program for Scientific Translations Ltd, 1962).

\bibitem{JucBan:77a}
A.P. Jucys and A.A. Bandzaitis, {\it Theory of Angular Momentum in Quantum Mechanics}, (Mokslas, Vilnius, 1977). [in Russian]

\bibitem{FanRac:59a}
U.~Fano and G.~Racah, {\it Irreducible Tensorial Sets}, (Academic Press, 1959).

\bibitem{Jud:67a}
B.R. Judd, {\it Second Quantization and Atomic Spectroscopy}, (The Johns Hopkins Press, Baltimore, MD, 1967).

\bibitem{RudKan:84a}
Z. Rudzikas and J. Kaniauskas, {\it Quasispin and Isospin in the Theory of Atom}, (Mokslas, Vilnius, 1984). [in Russian]

\bibitem{Gaigatal:85}
G. Gaigalas, J. G. Kaniauskas, and Z. Rudzikas, Diagrammatic technique of the angular momentum
theory and second quantization, Liet. Fiz. Rink. (English transl. - Sov. Phys. Coll.) {\bf 25}, 3-13 (1985). [in Russian]

\bibitem{Rud:97a}
Z. Rudzikas, {\it Theoretical Atomic Spectroscopy}, (Cambridge University Press, Cambridge, 1997).

\bibitem{Gaietal:2000a}
G. Gaigalas, S. Fritzsche, and Z. Rudzikas, Reduced coefficients of fractional parentage and matrix elements of the tensor $W^{(k_q k_j)}$ in $jj$-coupling, At. Data Nucl. Data Tables {\bf 76} (2), 235-269 (2000), https://doi.org/10.1006/adnd.2000.0844.

\bibitem{Meratal:85}
G. Merkelis, G. Gaigalas, and Z. Rudzikas, Irreducible tensorial form of the effective Hamiltonian
of an atom and the diagrammatic representation in the first two orders of the stationary
perturbation theory, Liet. Fiz. Rink. (English transl. - Sov. Phys. Coll.) {\bf 25}, 14-31 (1985). [in Russian]

\bibitem{Jonsson:23}
P. J\"onsson, G. Gaigalas, Ch. F. Fischer, J. Biero\'n, I.P. Grant, T. Brage, J. Ekman, M. Godefroid, J. Grumer, J. Li, and W. Li, GRASP manual for users, Atoms, {\bf 11}, 68 (2023). (https://doi.org/10.3390/atoms11040068)

\bibitem{Gusatal:2017}
S. Gustafsson, P. J\"onsson, C. Froese Fischer, and I. P. Grant, Combining multiconfiguration and perturbation methods: perturbative estimates of core-core electron correlation contributions to excitation energies in Mg-like iron, Atoms {\bf 5} (1), 3 (2017), https://doi.org/10.3390/atoms5010003.

\bibitem{Jonsson:17}
P. J\"onsson, G. Gaigalas, P. Rynkun, L. Rad\v{z}i\=ut\.e, J. Ekman, S. Gustafsson, H. Hartman, K. Wang, M. Godefroid, Ch.F. Fischer, I. Grant, T. Brage, and G.D. Zanna, Multiconfiguration Dirac-Hartree-Fock calculations with spectroscopic accuracy: applications to astrophysics, Atoms, {\bf 5}, 16 (2017), https://doi.org/10.3390/atoms5020016

\bibitem{Bogetal:1998} 
P. Bogdanovich, G. Gaigalas, and A. Momkauskait{\.e}, Accounting for correlation corrections
to interconfigurational matrix elements, Lith. J. Phys. {\bf 38} (5), 443-451 (1998). [in Russian]

\bibitem{Kar:96a}
R. Karazija, {\it Introduction to the Theory of X-Ray and Electronic Spectra of Free	Atoms}, (Plenum Press, New York, 1996).

\bibitem{Gaigalasetal:2020}
G. Gaigalas, P. Rynkun, L. Rad\v{z}i\={u}t\.{e}, D. Kato, M. Tanaka, and P. J\"onsson, Energy level structure and transition data of Er$^{2+}$, Astrophys. J Suppl. Ser. {\bf 248} (1), 13 (2020), https://doi.org/10.3847/1538-4365/ab881a.

\bibitem{Gaigalasetal:2021}
G. Gaigalas, P. Rynkun, L. Rad\v{z}i\={u}t\.{e}, P. J\"onsson, and K. Wang, Energy and transition data computations for P-like ions: As, Kr, Sr,, Zr, Mo, and W, At. Data Nucl. Data Tables {\bf 141} 101428 (2021), https://doi.org/10.1016/j.adt.2021.1.1428.

\bibitem{grasp2023}
Y.T. Li, K. Wang, R. Si, M. Godefroid, G. Gaigalas, Ch.Y. Chen, and P. J{\"o}nsson, Reducing the computational load - atomic multiconfiguration calculations based on configuration state function generators, Comput. Phys. Commun. {\bf 283}, 108562 (2023), https://doi.org/10.1016/j.cpc.2022.108562.

\bibitem{Bogdanovich&Kisielius2012}
P. Bogdanovich and R. Kisielius, Theoretical energy level spectra and transition data for 4p(6)4d,
   4p(6)4f and 4p(5)4d(2) configurations of W37+ ion, At. Data Nucl. Data Tables {\bf 98} (4), 557-565 (2012), https://doi.org/10.1016/j.adt.2011.11.004.

\bibitem{Bogdanovich&Kisielius2013}
P. Bogdanovich and R. Kisielius, Theoretical energy level spectra and transition data for 4p(6)4d(2),
   4p(6)4d4f, and 4p(5)4d(3) configurations of W36+, At. Data Nucl. Data Tables {\bf 99} (5), 580-594 (2013), https://doi.org/10.1016/j.adt.2012.11.001.

\bibitem{Karpuskiene&Kisielius2022}
R. Karpu{\v s}kien{\.e} and R. Kisielius, Theoretical level energies and transition data for 4p(6)4d(8),
   4p(5)4d(9) and 4p(6)4d(7)4f configurations of W30+ ion, At. Data Nucl. Data Tables {\bf 143}, 101478 (2022), https://doi.org/10.1016/j.adt.2021.101478.

\bibitem{Kisielius_etal2015}
R. Kisielius, V. P. Kulkarni, G. J. Ferland, P. Bogdanovich, D. Som, and M. L. Lykins, Atomic data for ZN II: 
   improving spectral diagnostics of chemical evolution in hihg-redshift galaxies, Astrophys. J {\bf 804} (1), 76 (2015), https://doi.org/10.1088/0004-637X/804/1/76.

\bibitem{Bog:2005} 
P. Bogdanovich, Modeern methods of multiconfiguration studies of many-electron highly charged ions, Nucl. Instrum. Methods Phys. Res. B {\bf 235} (1-4), 92-99 (2005), https://doi.org/10.1016/j.nimb.2005.03.152.

\bibitem{Bogetal:1999} 
P. Bogdanovich, R. Karpu{\v s}kien{\.e}, and Z. Rudzikas, Calculation of electronic transitions in S IX, Phys. Scr. {\bf T80B}, 474-475 (1999), https://doi.org/10.1238/Physica.Topical.080a00474. 

\bibitem{Karpuskiene_etal2013}
R. Karpu{\v s}kien{\.e}, P. Bogdanovich, and R. Kisielius, Significance of $M2$ and $E3$ transitions for $4p^{5}4d^{N+1}$- and $4p^{6}4d^{N-1}4f$-configuration metastable-level liftimes, Phys. Rev. A {\bf 88}, 022519 (2013), https://doi.org/10.1103/PhysRevA.88.022519.

\bibitem{Aggarwal_etal2016}
K.M. Aggarwal, P. Bogdanovich, F.P. Keenan, and R. Kisielius, Energy levels and radiative rates for Cr-like Cu VI and Zn VII, At. Data Nucl. Data Tables {\bf 111-112}, 280-345 (2016), https://doi.org/10.1016/j.adt.2016.03.001.

\bibitem{Karpetal:2004} 
R. Karpu{\v s}kien{\.e}, P. Bogdanovich, and A. Udris, {\it Ab initio} oscillator strengths and radiative lifetimes for Ca IX, J. Phys. B: At. Mol. Opt. Phys. {\bf 37} (10), 2067 (2004), https://doi.org/10.1088/0953-4075/37/10/006.

\bibitem{Bogetal:2001} 
P. Bogdanovich, R. Karpu{\v s}kien{\.e}, and I. Martinson, Theoretical calculation of transition probabilities and lifetimes of levels of the $4d^9 5p$ configuration of Cd III, Opt. Spectrosc. {\bf 90} (1), 4-7 (2001), https://doi.org/10.1134/1.1343537. 

\bibitem{Rynkun_2019}
P. Rynkun, G. Gaigalas, and P. J\"onsson, Theoretical investidation of energy levels and transition data for S II, Cl III, Ar IV, Astronomy \& Astrophysics {\bf 623}, A155 (2019), https://doi.org/10.1051/0004-6361/201834931.

\end{thebibliography}

\bibliographystyle{pccp}

\end{document}